\documentclass[10pt,a4paper,twoside,twocolumn,notitlepage]{article}

\usepackage{graphicx}% Include figure files
\usepackage{dcolumn}% Align table columns on decimal point
\usepackage{bm}% bold math

\usepackage[english]{babel}
\usepackage{amsmath}
\usepackage{amssymb}
\usepackage{units}
\usepackage{subfigure}
\usepackage{cite}
\usepackage{multirow}

\newdimen\Parindent\newdimen\Parskip
\Parindent=\parindent\Parskip=\parskip%save old values
\newcommand{\hb} {HERA-B\ }
\newcommand{\ks}{$K_{S}^{0}$}

\newcommand{\lam}{$\Lambda$}
\newcommand{\lamb}{$\bar{\Lambda}$}

\newcommand{\ksto}{$K_{S}^{0}\to \pi^+\pi^-$}
\newcommand{\lamto}{$\Lambda\to p\pi^-$}
\newcommand{\lambto}{$\bar{\Lambda}\to \bar{p}\pi^+$}

\newcommand{\XXX}[1]{}

\begin{document}

\titlepage{ %titlepage

\flushleft{
DESY-08-179\\
\today \\
}
\vspace{3mm}

\center{\LARGE\bf V$^0$ production in p+A collisions at \mbox{\boldmath{$\sqrt{s}\mathrm{~ = \unit[41.6]{GeV}}$}}}

\vspace{5mm}
\center{\large\bf The HERA-B Collaboration}\\
\vspace{2mm}

I.~Abt$^{24}$,
M.~Adams$^{11}$,
M.~Agari$^{14}$,
H.~Albrecht$^{13}$,
A.~Aleksandrov$^{30}$,
V.~Amaral$^{9}$,
A.~Amorim$^{9}$,
S.~J.~Aplin$^{13}$,
V.~Aushev$^{17}$,
Y.~Bagaturia$^{13,37}$
% ...

\vspace{5mm}
\noindent
% ...
$^{9}${\it LIP Coimbra, P-3004-516 Coimbra,  Portugal~$^{f}$} \\
% ...
$^{11}${\it Institut f\"ur Physik, Universit\"at Dortmund, D-44221 Dortmund, Germany~$^{d}$} \\
% ...
$^{13}${\it DESY, D-22603 Hamburg, Germany} \\
$^{14}${\it Max-Planck-Institut f\"ur Kernphysik, D-69117 Heidelberg, Germany~$^{d}$} \\
% ...
$^{17}${\it Institute for Nuclear Research, Ukrainian Academy of Science, 03680 Kiev, Ukraine~$^{h}$} \\
% ...
$^{24}${\it Max-Planck-Institut f\"ur Physik, Werner-Heisenberg-Institut, D-80805 M\"unchen, Germany~$^{d}$} \\
% ...
$^{30}${\it DESY, D-15738 Zeuthen, Germany} \\
% ...
$^{37}${\it visitor from High Energy Physics Institute, 380086 Tbilisi, Georgia} \\
% ...
$^{d}$ supported by the Bundesministerium f\"ur Bildung und Forschung, FRG, under contract numbers 05-7BU35I, 05-7DO55P, 05-HB1HRA, 05-HB1KHA, 05-HB1PEA, 05-HB1PSA, 05-HB1VHA, 05-HB9HRA, 05-7HD15I, 05-7MP25I, 05-7SI75I \\
% ...
$^{f}$ supported by the Portuguese Funda\c c\~ao para a Ci\^encia e Tecnologia under the program POCTI \\
% ...
$^{h}$ supported by the National Academy of Science and the Ministry of Education and Science of Ukraine \\
% ...

\abstract{ Inclusive doubly differential cross sections
  $d^2\sigma_{pA}/dx_Fdp_T^2$ as a function of Feynman-x ($x_F$) and
  transverse momentum ($p_T$) for the production of \ks, \lam\ and
  \lamb\ in proton-nucleus interactions at \unit[920]{GeV} are
  presented. The measurements were performed by HERA-B in the negative
  $x_F$ range ($-0.12<x_F<0.0$) and for transverse momenta up to
  $p_T=\unit[1.6]{GeV/c}$.  Results for three target materials:
  carbon, titanium and tungsten are given.  The ratios of production
  cross sections are presented and discussed.  The Cronin effect is
  clearly observed for all three $V^0$ species.  The atomic number
  dependence is parameterized as $\sigma_{pA} = \sigma_{pN} \cdot
  A^\alpha$ where $\sigma_{pN}$ is the proton-nucleon cross section.
  The measured values of $\alpha$ are all near one.  The results are
  compared with EPOS\,1.67 and PYTHIA\,6.3.  EPOS reproduces the data
  to within $\approx20\%$ except at very low transverse momentum.
}\\ %end of abstract

\vspace{3mm}

\noindent{\small{\it PACS} 13.85.Hd, 14.20.Jn, 14.40.Aq}

}
\twocolumn

\section{Introduction} 

The study of strange particle production in proton induced reactions
has a long history, starting from the discovery of strange particles
in cosmic rays in the 1950s.  Numerous studies have been made (see
[1-20, and references therein]) including fixed-target experiments at
Center-of-Mass (CM) energies up to 40\,GeV, mainly with bubble
chambers, as well as at CERN's Intersecting Storage Ring in the 1970s
and early 1980s~( see~\cite{busser76,erhan79,drij81,drij82}) and later
at the SPS Collider~\cite{ua1_96}. More recently, studies of
strangeness production at a CM energy of $\mathrm{200\,\mathrm{GeV}}$
in both proton-proton and deuteron-gold collisions at RHIC have been
published~\cite{star_dAu, star_pp}.  A detailed understanding of the
underlying production mechanism, particularly in proton-nucleus
interactions, is lacking. Further work, both experimental and
theoretical, is needed both to improve the modeling of atmospheric
cosmic ray showers, and to serve as a reference for strangeness
production studies in heavy ion collisions. The study presented in
this paper was performed at the highest available fixed-target energy
and benefits from a large sample size.

We present the doubly differential cross sections for \ks , \lam , and
\lamb\ production in proton collisions with carbon, titanium and
tungsten targets at a CM energy of $\sqrt{s} = 41.6\,\mathrm{GeV}$ as
a function of the squared transverse momentum range ($p_T$) in the
range ($0 < p_T^2 < 2.5\mathrm{(GeV/c)^2}$) and Feynman-x ($x_F$) in
the range ($-0.12 < x_F < 0.0$). The cross sections and derived
quantities are compared to predictions obtained from
PYTHIA~6.3\cite{pyt_63} and the EPOS~1.67 event
generator~\cite{EPOS_06}.  PYTHIA is not designed to model
proton-nucleus interactions, but the comparison is nonetheless
instructive. EPOS is an
event model currently under development which has recently been shown
to accurately account for many features of proton-proton\cite{star_pp}
and deuteron-gold collisions at RHIC\cite{EPOS_06}.  The EPOS model is
based on parton-parton interactions in which cascades of usually
off-shell partons (``parton ladders'') are produced which eventually
hadronize into the observed final state hadrons. More than one parton
ladder is generally produced. In the case of proton-nucleus
collisions, the partons representing the ladder rungs can
``rescatter'' with other target nucleons via elastic or inelastic
interactions. This leads to increased screening and also $p_T$
broadening with increasing target mass number.

In the following sections we briefly describe the detector, the
analysis and finally present the results. The study presented here
supersedes the previously published HERA-B study~\cite{v0paper} whose
results are inconsistent with those presented herein largely due to
errors in the detector description used for the previous study.

\section{\hb experiment and data sample}

\hb was a fixed target experiment at the proton storage ring of
HERA at DESY~\cite{har95}.  Collisions were produced by inserting one
or more wire targets into the halo of the \unit[920]{GeV/c} proton beam. The
center-of-mass energy in the proton-nucleon system was $\sqrt{s} =
\unit[41.6]{GeV}$.

The detector was designed and built as a magnetic spectrometer with a
forward acceptance of 15-220 mrad in the bending (horizontal) and
15-160 mrad in the non-bending (vertical) plane.  The target
system~\cite{ehr04} consisted of two stations separated by about 5~cm
with four wires each.  The wires were positioned above, below, and on
either side of the beam and were made of various materials including
carbon, titanium and tungsten.  The vertex detector system
(VDS)~\cite{kno03} was a planar micro-strip vertex detector providing
a precise measurement of primary and secondary vertices. The VDS
consisted of 8 stations (with 4 stereo views each) of double-sided
silicon strip detectors mounted in movable Roman Pots which allowed
operation as near as 10 mm from the beam and provided for retraction
during beam manipulations.  The vacuum vessel housing the detector
was an integral part of the HERA proton ring.  The VDS was
followed by a large aperture dipole magnet with a field integral of
$\unit[2.13]{Tm}$, and a set of tracking chambers (OTR)~\cite{hoh01}
consisting of $\approx$95,000 channels of honeycomb drift cells.
Particle identification was performed by a Ring Imaging Cherenkov
detector~\cite{ari04}, an electromagnetic calorimeter~\cite{avo01}
and a muon system~\cite{eig01}.

This analysis is based on about $10^7$ interactions on each of carbon,
titanium and tungsten targets. The data set is a subsample of the full
minimum bias data set ($2 \cdot 10^8$ events) which was taken over a
three-day period from a single filling of protons in the HERA proton
ring to minimize systematic uncertainties. Only one of the three
target wires was in use at a time.  All data were recorded with an
interaction rate of 1.5~MHz, corresponding to about one inelastic
interaction per six bunch crossings.  Non-empty events were selected
using an interaction trigger which required at least 20 hits in the
RICH detector (compared to an average of 33 for a full ring from a
$\beta=1$ particle~\cite{ari04}) or an energy deposit of at least
1~GeV in the electromagnetic calorimeter. The trigger was sensitive to
more than 97\% of the total inelastic cross section
$\sigma_{inel}$~\cite{bru07}.  The data sample also includes about
$5\cdot 10^5$ events per target selected at random, with no trigger
requirement, which were taken at a 10~Hz rate throughout the data
taking period. These ``random'' events were used for luminosity
determination and systematic studies.

The entire $V^0$ candidate reconstruction chain was based exclusively
on information from the VDS and OTR.  All events were reconstructed
with the standard \hb analysis package~\cite{abt02}.

\section{Data analysis}
The \ks, \lam\ and \lamb\ particles are reconstructed from their two
particle decays \ksto, \lamto\ and \lambto, respectively.

For this analysis, a track consists of matched reconstructed OTR and
VDS track segments. A search for a primary vertex is performed using
them and, if successful, the interaction point is taken to be the
location of the found vertex.  If unsuccessful, the position of the
target wire together with the average position of interactions along
the wire are used.  In each event, a full combinatorial search for
$V^0$ candidates is then performed.

$V^0$ candidates are selected from all pairs of oppositely charged
tracks which form a secondary vertex downstream of the interaction
point.  The minimum distance between the two tracks of a pair is
required to be less then 0.14\,cm.  The $\pi^+\pi^-$, $p\pi^-$ and
$\bar{p}\pi^+$ mass hypotheses are assigned in turn. If the
$\pi^+\pi^-$ invariant mass hypothesis lies in the region
$0.44<M_{\pi^+\pi^-}<0.56\,\mathrm{GeV/c^2}$ or either the $p\pi^-$ or
$\bar{p}\pi^+$ invariant mass hypothesis lie in the region
$1.09<M_{p\pi^-/\bar{p}\pi^+}<1.14\,\mathrm{GeV/c^2}$, the pair is accepted
for further analysis.  To reduce cross-contamination of \ks's and
\lam/\lamb\ samples, pairs with $\pi^+\pi^-$ invariant mass
in the range $0.476<M_{\pi^+\pi^-}<0.515\,\mathrm{GeV/c^2}$ 
are excluded from the \lam\ and \lamb\ analyses, and pairs
with
$p\pi^-$ and $\bar{p}\pi^+$ mass hypotheses in the range
$1.109<M_{p\pi^-/\bar{p}\pi^+}<1.121\,\mathrm{GeV/c^2}$ are excluded from
the \ks\ analysis.  

Finally, a cut on the product of the transverse momenta of the decay
products relative to the flight direction of the $V^{0}$ candidate
and the proper decay length of the $V^{0}$, $\tilde{p_T} \cdot c\tau
>0.05 \ \mathrm{GeV/c \cdot cm}$, is applied.  This requirement
rejects short-lived combinatorial background from the target region
and also reduces background from $\gamma \to e^+e^-$ conversions.

\begin{figure*}
  \begin{center}
    \subfigure[]{
      \includegraphics[clip,width=0.3133\textwidth]
      {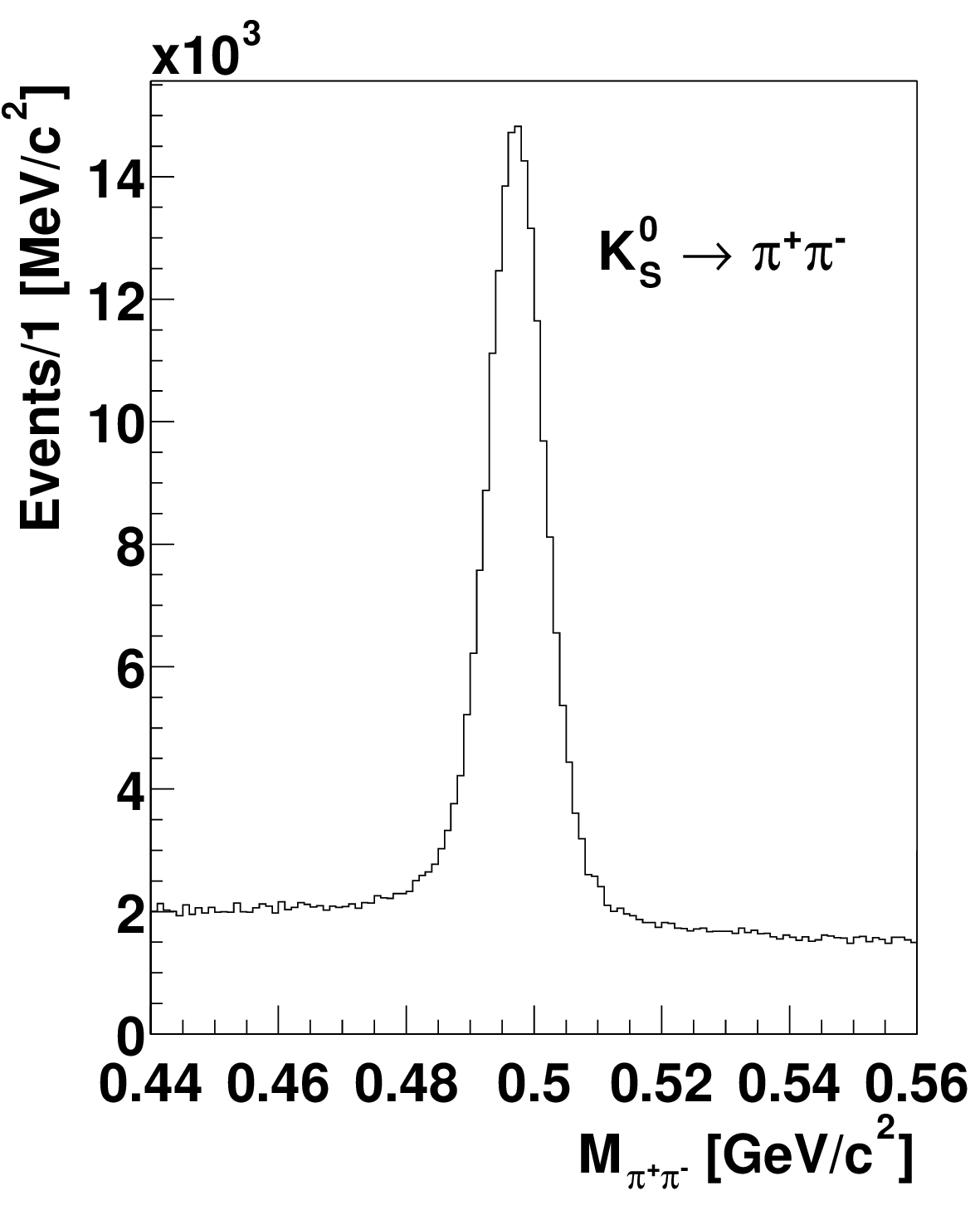}}     
    \subfigure[]{
      \includegraphics[clip,width=0.3133\textwidth]
      {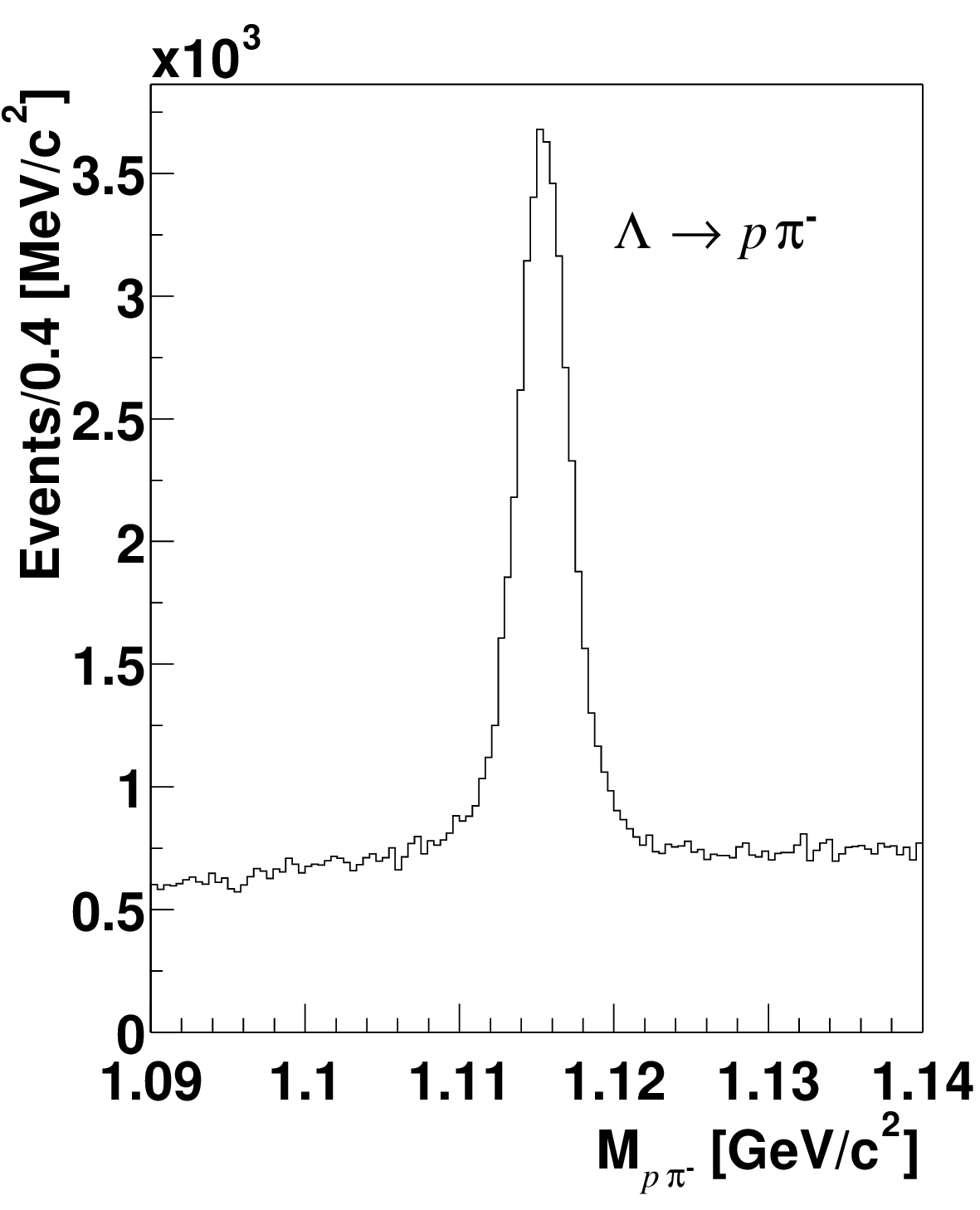}}
    \subfigure[]{
      \includegraphics[clip,width=0.3133\textwidth]
      {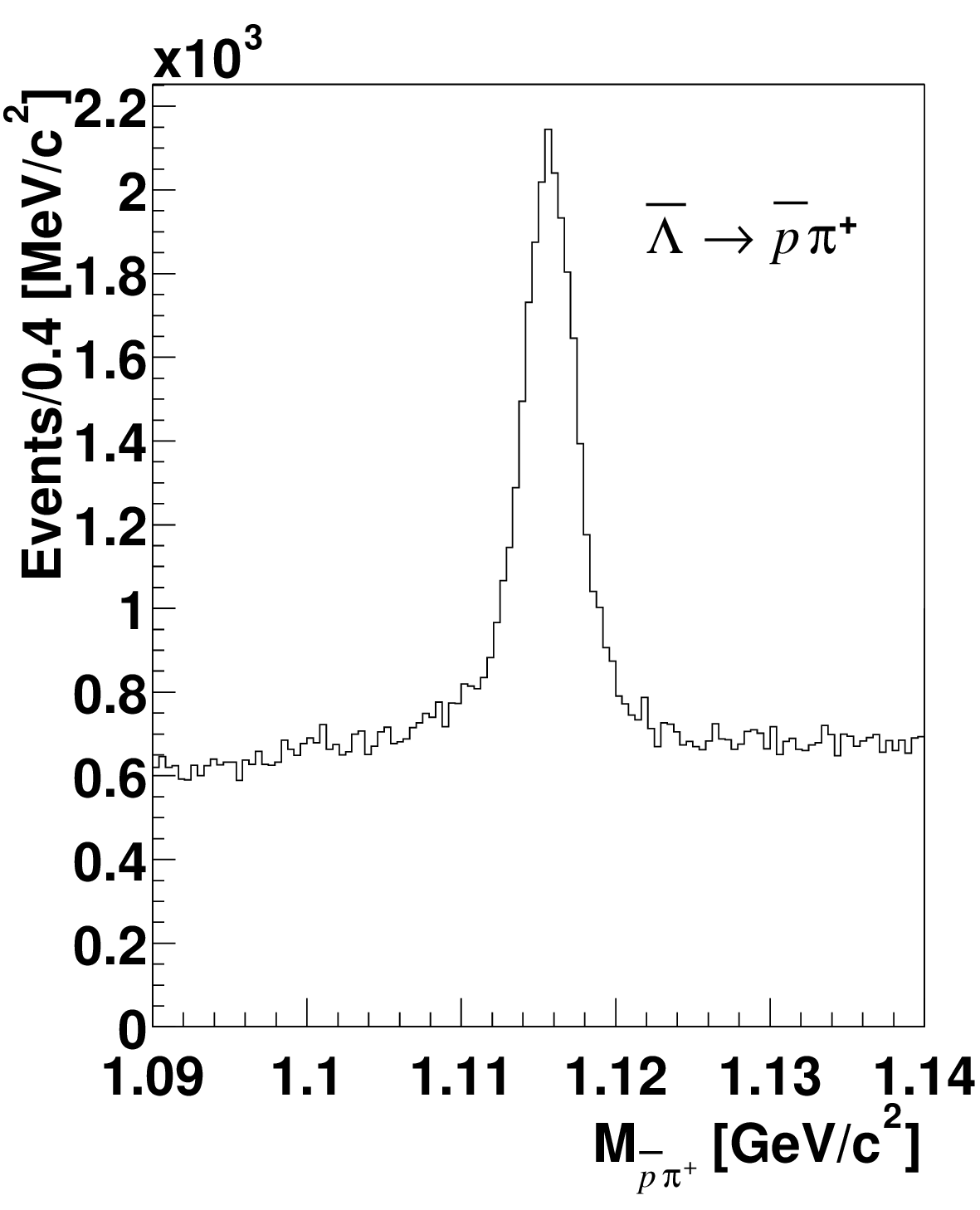}}
  \end{center}
\caption{The invariant mass distributions for oppositely charged 
  particle pairs, assuming a) $\pi^+\pi^-$, b) $p\pi^-$ and
  c) $\bar{p}\pi^+$ mass assignments for the carbon target sample after
  application of the selection criteria described in the text.}
\label{fig:InvMass}
\end{figure*}

The final invariant mass distributions for selected \ks, \lam, and
\lamb\ candidates from the carbon target sample are shown in Fig.
\ref{fig:InvMass}. Distributions from the other samples are similar.
The signals are clearly seen above a smooth background.  The yields of
$V^0$ are calculated from the number of entries in each bin of the
signal region within a $\pm 4\sigma$ window around the peak position
minus the background, which is taken from the left and right sidebands
with a width of $4\sigma$ each. A fit to the mass spectra using two
Gaussians with a common mean to describe the signal and a first order polynomial
to describe the background gives central mass values of 497.0\,MeV, 1115.3\,MeV and
1115.9\,MeV for \ks, \lam, and \lamb, respectively; all well within 1\,MeV of
the current PDG values~\cite{pdg08}.

\begin{table}
  \caption{The number of inelastic events ($N_{evt}$), number of reconstructed $V^0$ ($N_{V^0}$)
and the integrated luminosities~\cite{bru07} $\cal{L}_A$ in $\mathrm{mb^{-1}}$ for the indicated targets.}
  \label{tab:Statistics}
  \begin{center}
    \begin{tabular}{l @{\hspace{2mm}}c | @{\hspace{2mm}}c | @{\hspace{2mm}}c} 
      \hline\noalign{\smallskip}
      & \multicolumn{1}{c}{C} & \multicolumn{1}{c}{Ti} & \multicolumn{1}{c}{W}\\
      \hline\noalign{\smallskip}
      $N_{evt}$ & 9350000 & 9790000 & 10900000 \\
      \hline\noalign{\smallskip}
      $N_{K_{S}^{0}}$ & $152260\pm550$ & $210780\pm780$ & $265800\pm860$ \\
      \hline\noalign{\smallskip}
      $N_{\Lambda}$ & $30800\pm270$ & $45170\pm350$ & $65170\pm440$ \\
      \hline\noalign{\smallskip}
      $N_{\bar{\Lambda}}$ & $15220\pm240$ & $20990\pm310$ & $28840\pm430$ \\
      \hline\noalign{\smallskip}
      $\cal{L}_A$ & $40900.\pm1600.$ & $14880.\pm520.$ & $6110.\pm200.$ \\ 
      \hline\noalign{\smallskip}
    \end{tabular}
  \end{center}
\end{table}

The number of inelastic events, the signal yields obtained from the
selection described above, and the luminosity values~\cite{bru07} are summarized in
Table \ref{tab:Statistics} for each target material.

\section{Acceptance and visible kinematic region}
\label{sec:acceptance}

The reconstruction efficiencies for \ks, \lam\ and \lamb\ in the
selected decay channels are determined from Monte Carlo (MC) using the
FRITIOF 7.02 package \cite{pi92} for event generation.  FRITIOF is a
proton-proton, proton-nucleus and nucleus-nucleus collision generator
based on a model in which hadrons are treated as strings. 
The generated events are propagated through
the detector using the GEANT~3.21 package \cite{gea93}. Realistic
detector efficiencies, electronic noise and dead channel maps are
included in the simulation.  The MC events are processed through the
same reconstruction chain as the data.  The sizes of the MC samples used
for the efficiency calculations are about the same as those of the
data.  The uncertainties due to MC statistics are added in quadrature
with the statistical uncertainties of the data.

The total efficiency which includes geometric acceptance, track
reconstruction efficiency, and the efficiency of selection cuts,
depends on the kinematic variables and is, on average, 9\% for \ks,
and 5\% for \lam\ and \lamb\ inside the ``visible region'', defined as
$-0.12<x_F<0.0$ and $p_{T}^{2}<\unit[2.5]{GeV^2/c^2}$, for all $V^0$
types.  The efficiencies are determined on a grid in $x_{F}$ and
$p_{T}^{2}$ with 6 equal bins in $x_F$ and 10 equal bins in $p_T^2$
over the range given above, for a total of 60 bins. The grid-based
acceptance correction has the advantage of minimizing any biases due
to inaccuracies in the generated kinematic distributions.

For \ks\ mesons, the low $p_T$ bins of the lowest $x_F$ regions are
poorly populated due to low acceptance, and are therefore excluded.
Specifically, for a bin to be considered, we require that it contain
at least 10 events in both MC and data samples.  For \ks\ mesons, the
$x_F/p_T^2$ interval $[-0.12,-0.08]$ / $[0.0,0.5]~\mathrm{GeV/c^2}$
for all samples and in addition the interval $[-0.12,-0.10]$ /
$[0.5,0.75]$ $\mathrm{GeV/c^2}$ for the titanium sample are
excluded. For the total cross section, A-dependence, and
production-ratio studies, the data are summed either in slices of
$x_F$ or $p_T^2$. The results are limited to the kinematic range over
which all bins are populated.  Thus, for the
\ks , only the $x_F$ interval $-0.08<x_F<0.0$ is considered for such
studies.

Based on a MC study, a small correction is applied to account
for those $V^0$ particles which are produced in interactions with
the detector material. The corrections obtained reduce the acceptance
by $0.9\%$--$1.2\%$ for \ks, $1.0\%$--$1.4\%$
for \lam\ and $0.3\%$ for \lamb, depending on target material.

\begin{figure*}
  \newcommand{\sssz}{0.29}
  \begin{center}
    \subfigure[$p+C \rightarrow K^0_S + X$]{
      \includegraphics[clip,width=\sssz\textwidth]
      {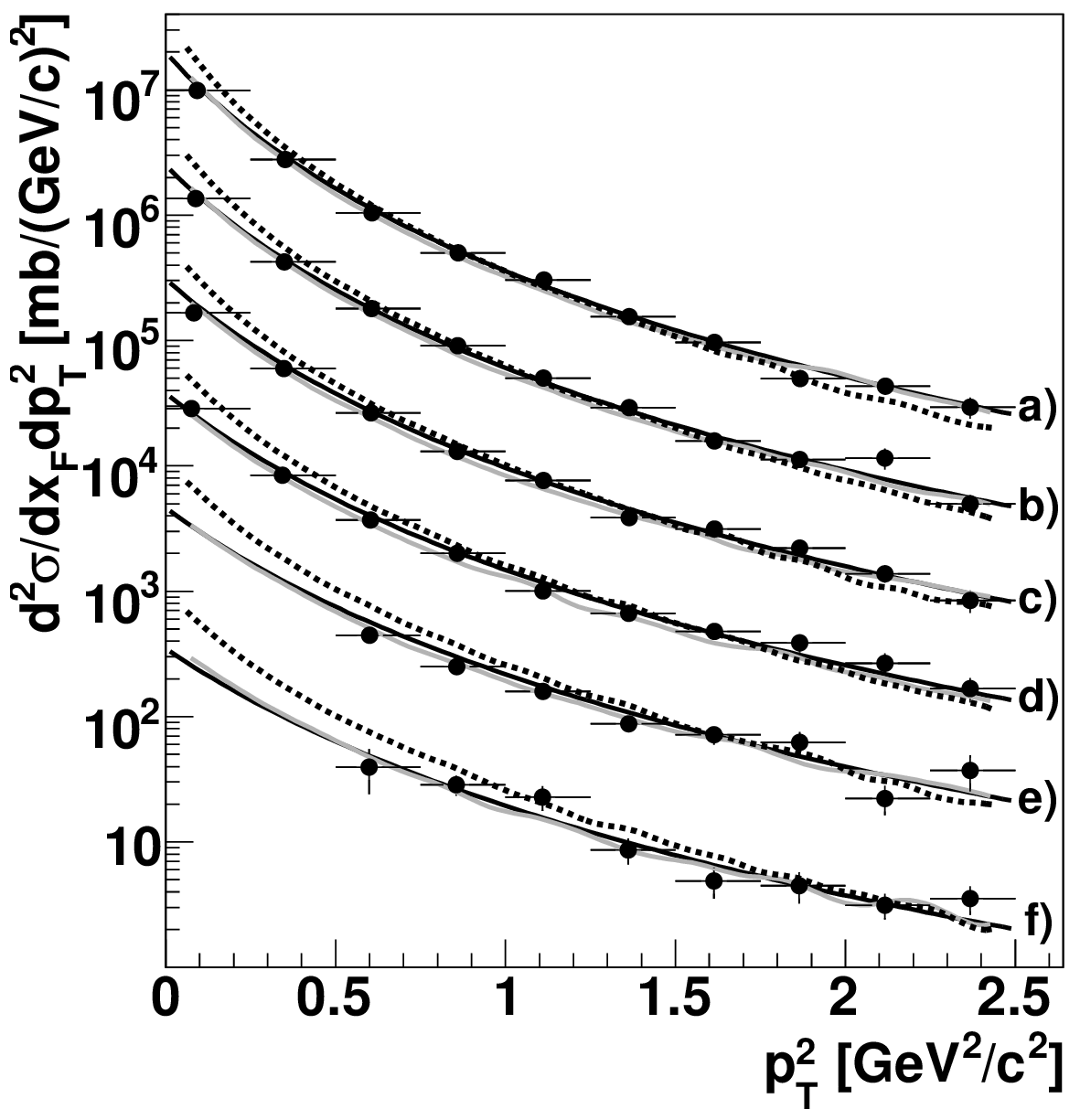}}
    \subfigure[$p+C \rightarrow \Lambda + X$]{
      \includegraphics[clip,width=\sssz\textwidth]
      {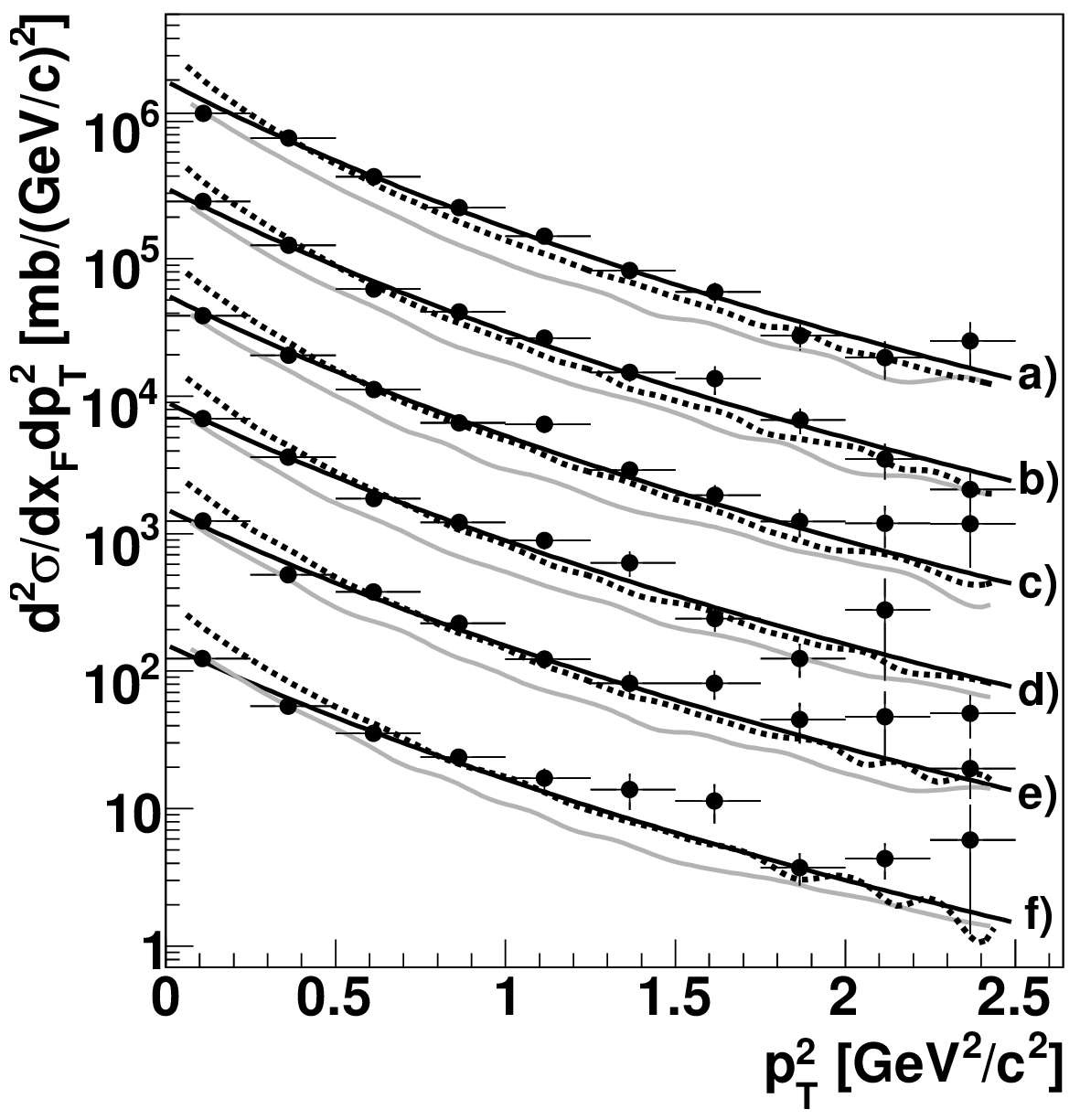}}
    \subfigure[$p+C \rightarrow \bar{\Lambda} + X$]{
      \includegraphics[clip,width=\sssz\textwidth]
      {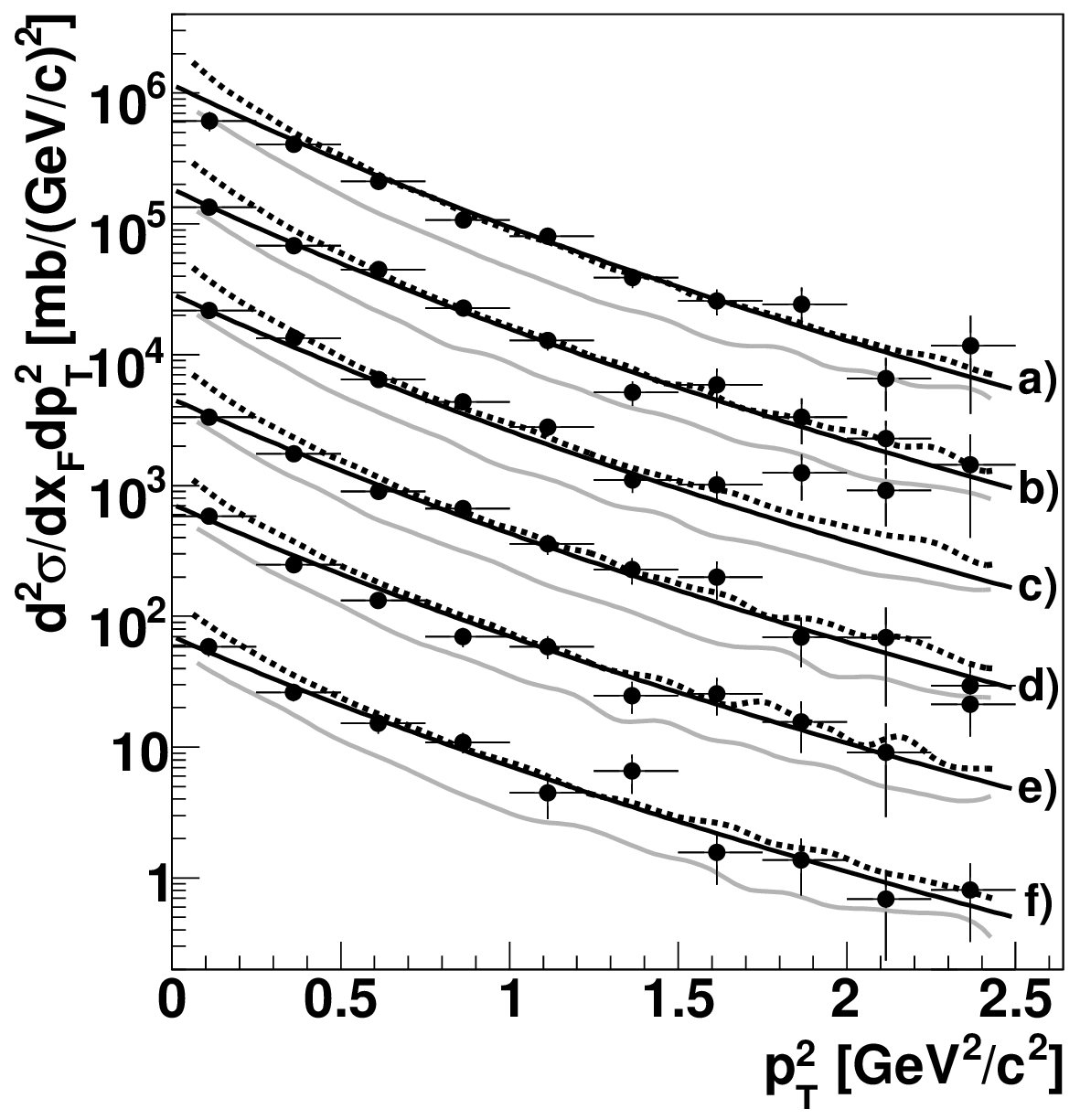}}
  \end{center}
  \begin{center}
    \subfigure[$p+Ti \rightarrow K^0_S + X$]{
      \includegraphics[clip,width=\sssz\textwidth]
      {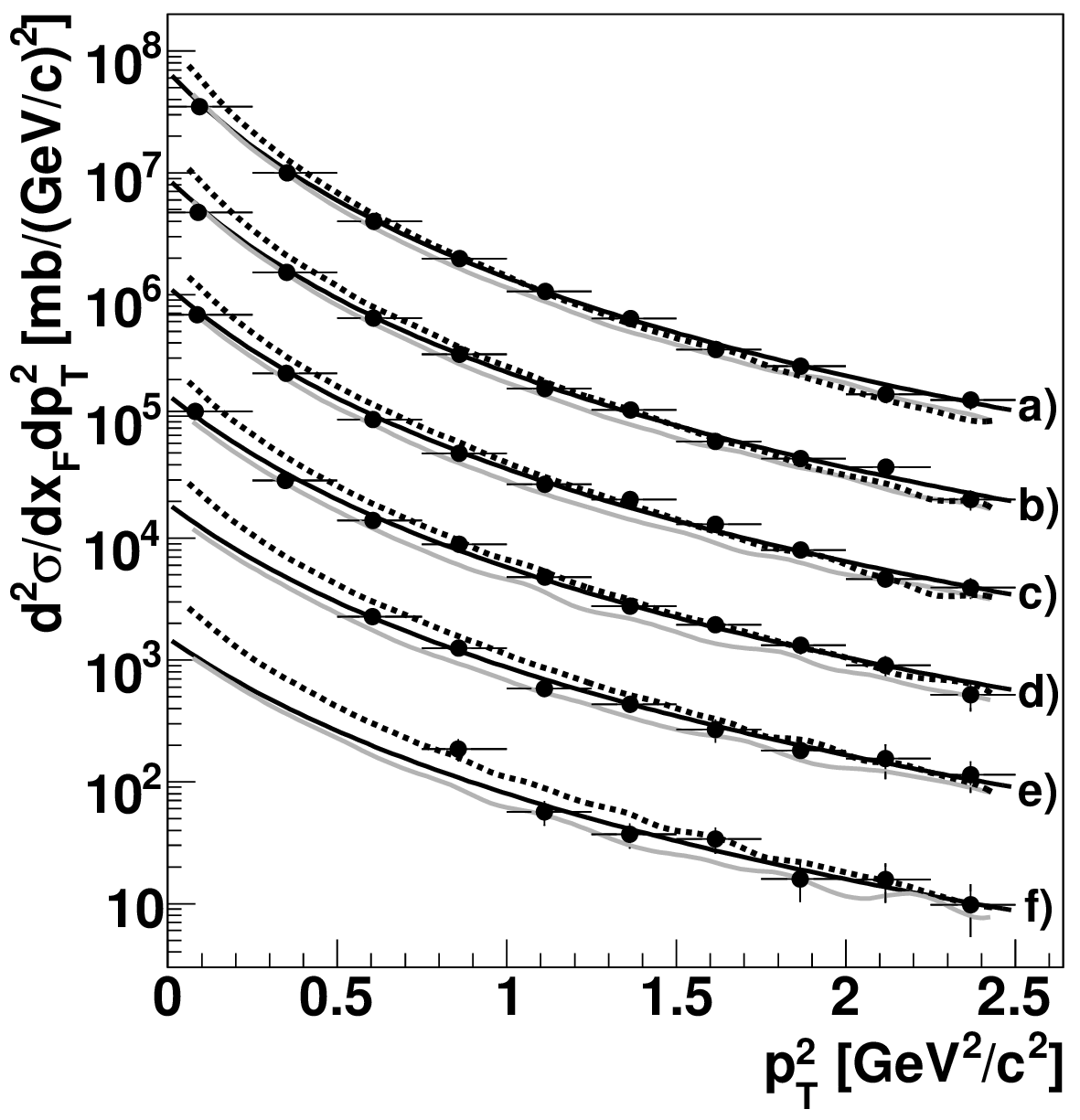}}
    \subfigure[$p+Ti \rightarrow \Lambda + X$]{
      \includegraphics[clip,width=\sssz\textwidth]
      {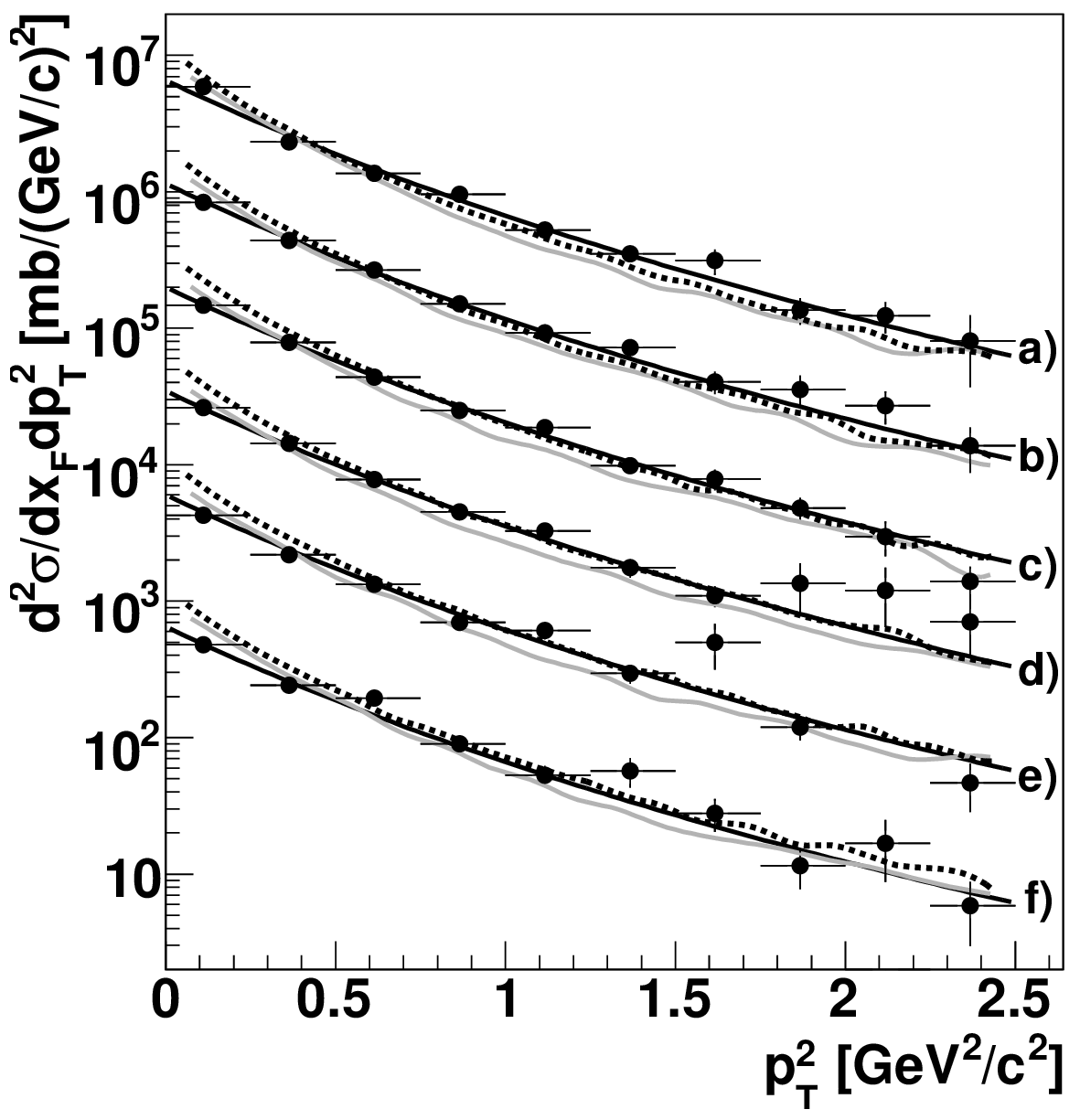}}
    \subfigure[$p+Ti \rightarrow \bar{\Lambda} + X$]{
      \includegraphics[clip,width=\sssz\textwidth]
      {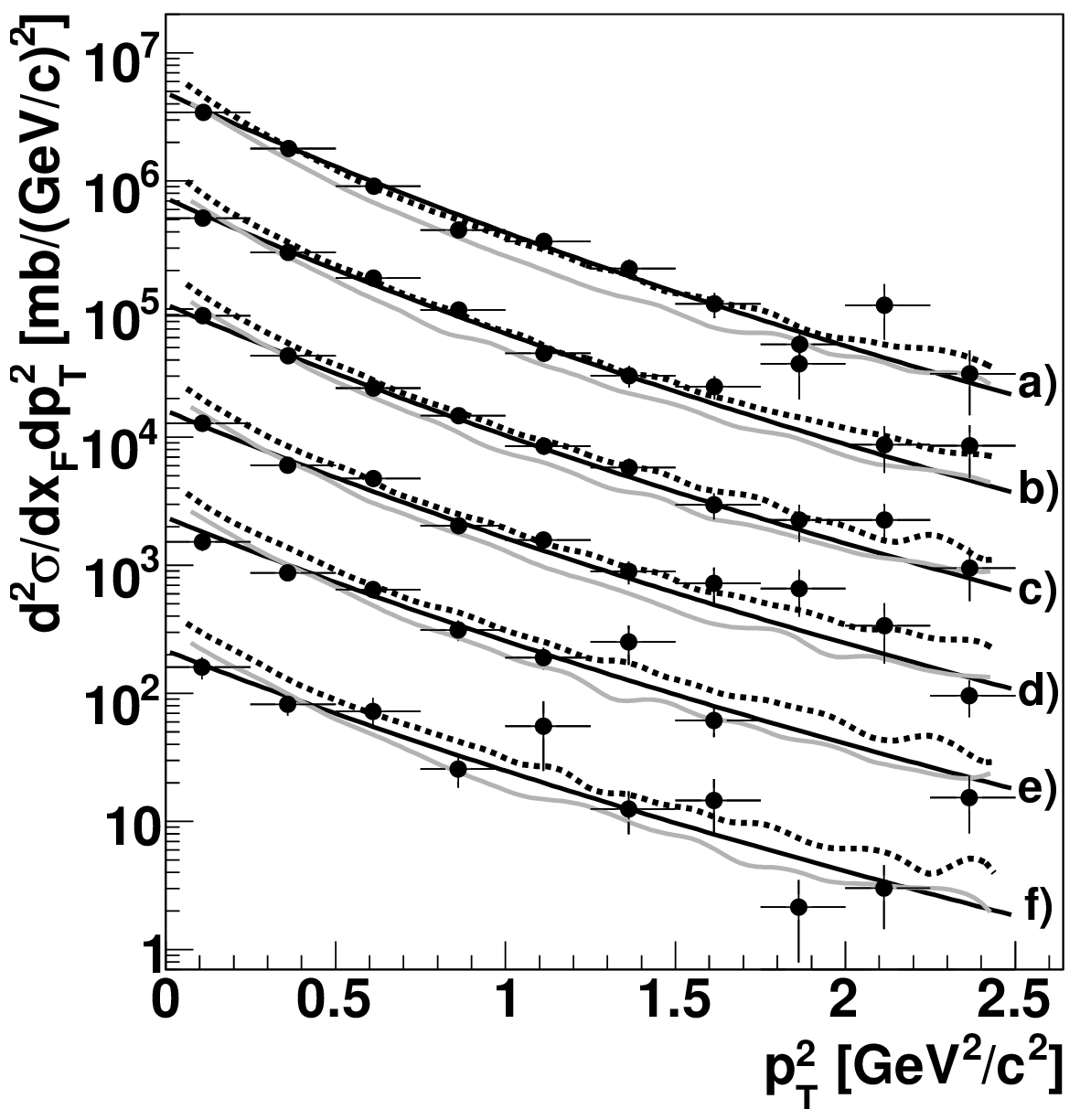}}
  \end{center}
  \begin{center}
    \subfigure[$p+W \rightarrow K^0_S + X$]{
      \includegraphics[clip,width=\sssz\textwidth]
      {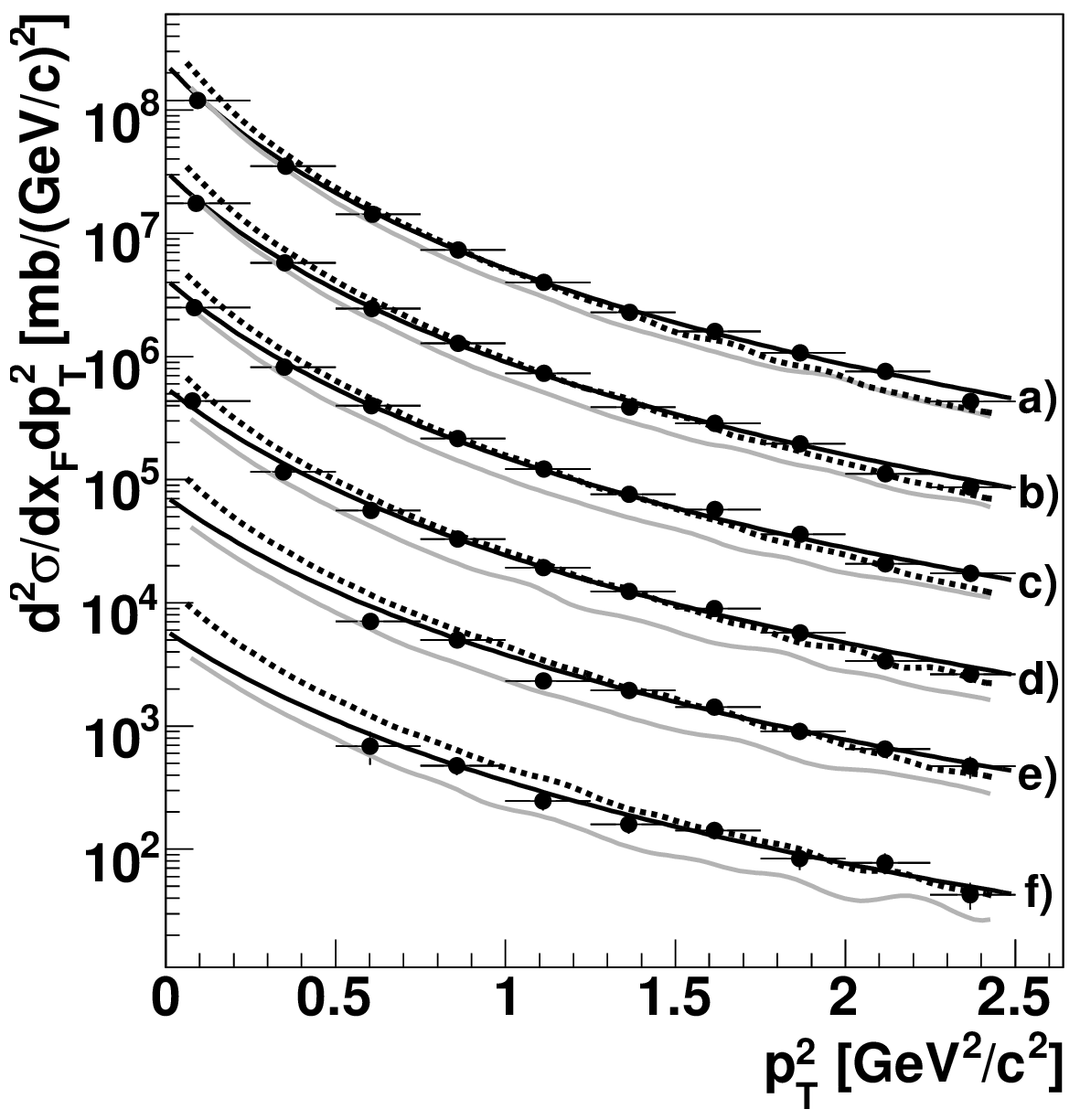}}
    \subfigure[$p+Ti \rightarrow \Lambda + X$]{
      \includegraphics[clip,width=\sssz\textwidth]
      {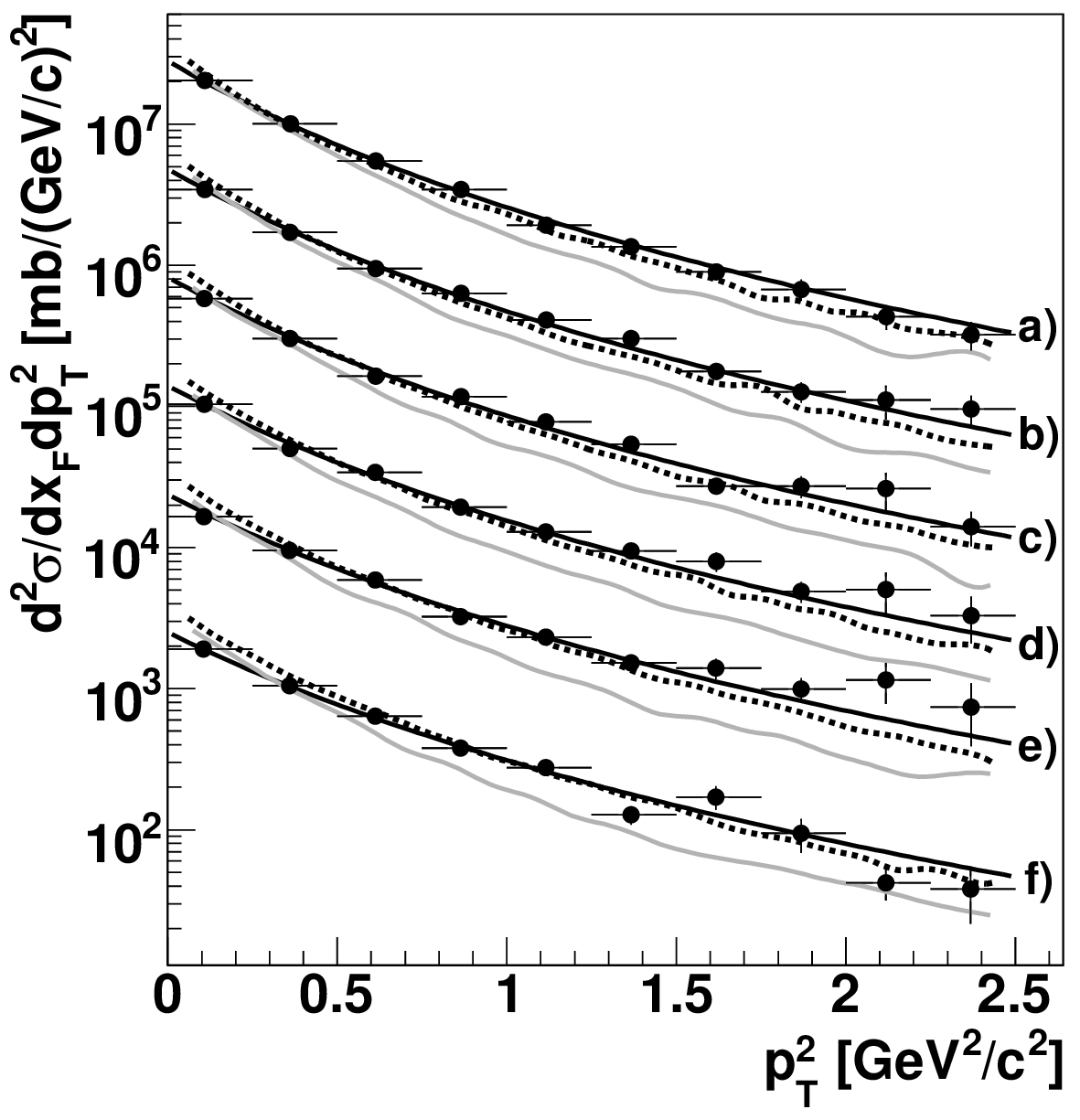}}
    \subfigure[$p+W \rightarrow \bar{\Lambda} + X$]{
      \includegraphics[clip,width=\sssz\textwidth]
      {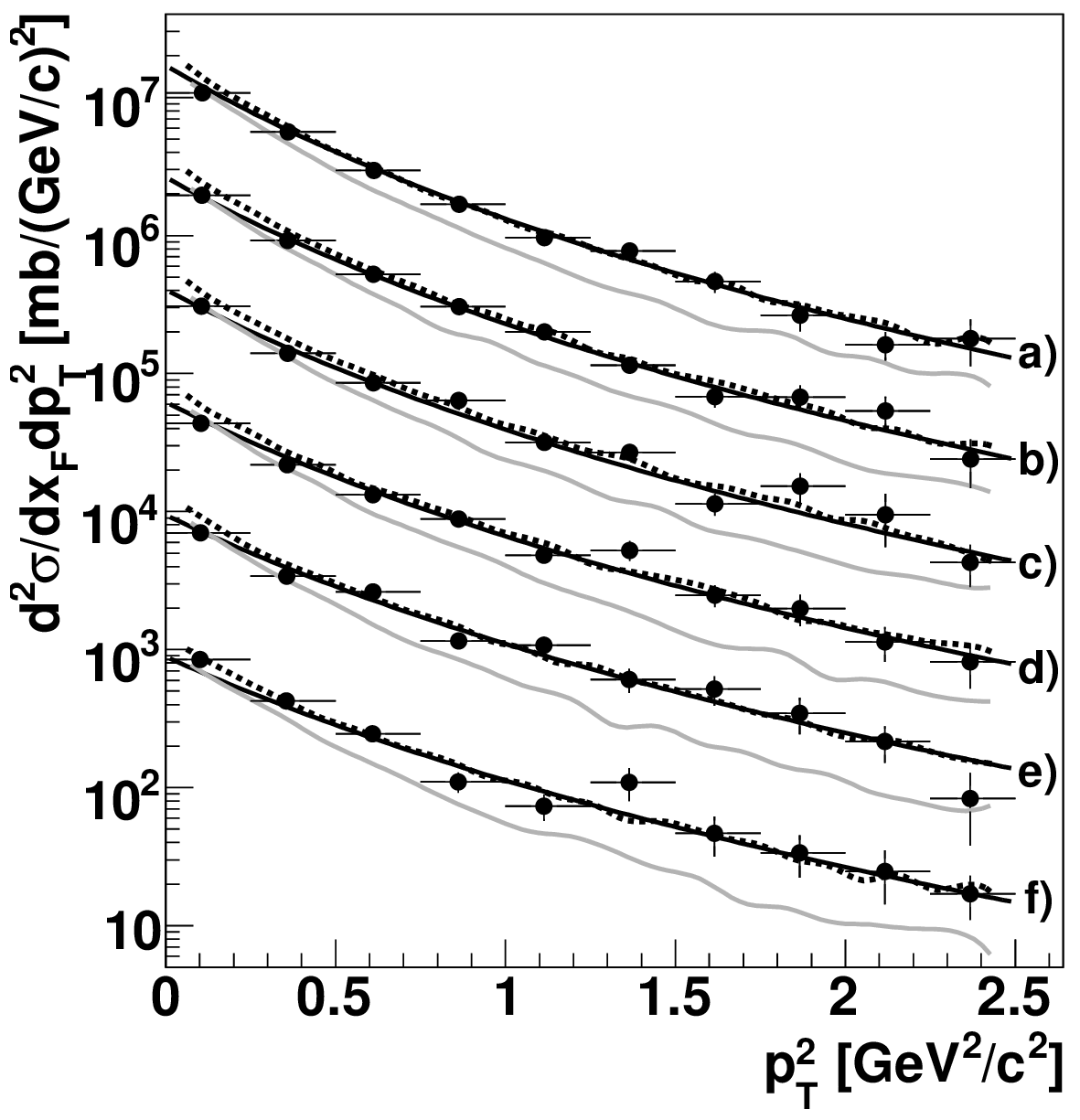}}
  \end{center}
  \caption{
    The measured inclusive doubly differential cross section
    $d^2\sigma_{pA}/dx_Fdp_T^2$ vs. $p_T^2$ in 6 $x_F$ slices for
    \ks , \lam\ , and \lamb\ production on carbon, titanium and
    tungsten targets. The error bars indicate the statistical
    uncertainties only. For display purposes, the cross sections in
    each $x_F$ slice have been multiplied by the following numbers
    (the letters correspond to those on the right of each curve):
    a)\,5000, b)\,1000, c)\,200, d)\,40, e)\,8, f)\,1. The $x_F$ ranges for each
    curve correspond to those given in Tables~\ref{tab:CS_K0s},
    \ref{tab:CS_Lam} and~\ref{tab:CS_Lam_bar}: a)\,-0.02 -- 0, b)\,-0.04
    -- -0.02, c)\,-0.06 -- -0.04, d)\,-0.08 -- -0.06, e)\,-0.10 -- -0.08,
    f)\,-0.12 -- -0.10.  The parameterizations discussed in the text are
    shown as dark solid lines. The light solid lines show the results
    of PYTHIA normalized to the ($x_F, p^2_T)$ bin (-0.01, 0.125
    $\mathrm{(GeV/c)^2}$) (separately for each plot). EPOS results are
    indicated by dashed lines. }
  \label{fig:DoubleCS}
\end{figure*}

\section{Experimental results}

The main results of this paper, the doubly differential cross
sections, are discussed in the following section. The subsequent
sections are devoted to discussions of quantities derived from these
numbers, such as A-dependence and production ratios.

\subsection{Doubly differential cross sections \label{sec:DoubleDiffCS}}

\begin{table*}
  \begin{center}

  \caption{
    Results of the combined power-law fits
    (Eq.~\ref{eq:Comb_xF_pT2_fit_func}) for the doubly differential cross
    sections $d^2\sigma_{pA}/dx_Fdp_T^2$.  Data were fitted in the
    acceptance region ($-0.12<x_F<0.0$ and $0.0
    <p_{T}^{2}<\unit[2.5]{GeV^2/c^2}$). Systematic uncertainties were
    not included in the fit and empty cells were excluded.  }

  \label{tab:Fit_Par}
  \begin{tabular}{l c | c | c | c | c | c }
    \hline\noalign{\smallskip}
    & $C_{0}$ [$\mathrm{mb/(GeV/c)^2}$]& $\beta$ & $A,{\ [\mathrm{GeV^{2}/c^2}]}$ & $B,{\ [\mathrm{GeV^{2}/c^2}]}$
    & $\text{n}$ & $\chi^2/{\mbox{\tiny DOF}}$\\
    \hline\noalign{\smallskip}
    & \multicolumn{6}{c}{$p + A \to K_{S}^{0} + X$}\\
    \hline\noalign{\smallskip}
    $C$ & $4893.\pm79.$ & $3.93\pm0.09$ & $0.53\pm0.02$ & $3.57\pm0.32$ & $22.85\pm0.64$ & $81./51$ \\
$Ti$ & $16650.\pm270.$ & $3.69\pm0.08$ & $0.50\pm0.02$ & $2.89\pm0.32$ & $20.71\pm0.70$ & $40./50$ \\
$W$ & $56980.\pm770.$ & $3.53\pm0.06$ & $0.48\pm0.01$ & $3.12\pm0.28$ & $19.53\pm0.58$ & $91./51$ \\

    \hline\noalign{\smallskip}
    & \multicolumn{6}{c}{$p + A \to \Lambda + X$}\\
    \hline\noalign{\smallskip}
    $C$ & $425.\pm21.$ & $6.90\pm0.96$ & $2.33\pm0.43$ & $2.4\pm1.4$ & $8.64\pm0.67$ & $91./55$ \\
$Ti$ & $1402.\pm52.$ & $6.40\pm0.87$ & $2.33\pm0.41$ & $  0  $ & $6.62\pm0.35$ & $59./54$ \\
$W$ & $6040.\pm260.$ & $3.98\pm0.31$ & $1.19\pm0.15$ & $2.34\pm0.63$ & $7.57\pm0.56$ & $62./55$ \\

    \hline\noalign{\smallskip}
    & \multicolumn{6}{c}{$p + A \to \bar{\Lambda} + X$}\\
    \hline\noalign{\smallskip}
    $C$ & $259.\pm19.$ & $9.9\pm2.6$ & $3.4\pm1.2$ & $3.8\pm3.2$ & $11.3\pm1.0$ & $72./54$ \\
$Ti$ & $1121.\pm82.$ & $11.0\pm3.2$ & $3.8\pm1.6$ & $7.0\pm4.6$ & $14.2\pm1.2$ & $63./52$ \\
$W$ & $3860.\pm270.$ & $4.86\pm0.59$ & $1.40\pm0.26$ & $4.4\pm1.4$ & $12.6\pm1.0$ & $59./55$ \\

    \hline\noalign{\smallskip}
  \end{tabular}
  \end{center}
\end{table*}

The doubly differential cross section for the state $V^0$ in the
$(i,j)$th bin of $(x_F, p^2_T)$ is computed from the following formula:
\begin{equation}
  \label{eq:DoubleDiffCS}
  \frac {d^2\sigma_{pA}^{V^0}(i,j)}{dx_Fdp_T^2} = \frac{N^{V^0}_{i,j}}
  {Br(V^0)\cdot{\cal L}_A\cdot \epsilon^{V^0}_{i,j} \cdot \Delta x_F\cdot{\Delta p_{T}^{2}}}\, ,
\end{equation}
where $Br(V^0)$ \cite{pdg08} is the branching ratio of the detected
decay and ${\cal L}_A$ is the integrated luminosity of the data set
for the specified target material (see Table~\ref{tab:Statistics}).
$N^{V^0}_{i,j}$ is the background-subtracted number of reconstructed
$V^0$ candidates in the $(i,j)$th bin of $(x_F,p_{T}^{2})$ and
$\epsilon^{V^0}_{i,j}$ is the corresponding efficiency calculated from
the MC as described in Sect.~\ref{sec:acceptance}. The bin widths
are 0.02 in $x_F$ and $\unit[0.25]{GeV/c^2}$ in $p_T$.

The values of the inclusive doubly differential cross sections,
$d^2\sigma/dx_Fdp_T^2$ for the full visible region are reported in
Tables~\ref{tab:CS_K0s}, \ref{tab:CS_Lam} and \ref{tab:CS_Lam_bar} for
all three target materials and illustrated in Fig.~\ref{fig:DoubleCS}.
The measurement resolutions in $x_F$ and $p_T^2$ are small compared to
the bin width. A discussion of systematic uncertainties can be found
in Sect.~\ref{sect:syst}. For the excluded bins (see
Sect.~\ref{sec:acceptance}), the values reported in the tables were
extrapolated using the fits described below.

The measured cross section distributions have the same general
behavior for all $V^0$ particles and can be described by the following
parameterization:
\begin{equation}
  \label{eq:Comb_xF_pT2_fit_func}
  \frac{d^2\sigma}{dx_Fdp_T^2} = C_0 \cdot  \left(1- \lvert x_F\rvert \right)^{n} 
  \cdot \left(1 + \frac{p_T^2}{A + B \cdot \lvert x_F\rvert} \right)^{-\beta}.
\end{equation}
The power law parameterization in $x_F$ is often used, particularly in
the fragmention region where the measured power has been used to
distinguish fragmentation models\cite{brodsky}. While the
parametrization has no theoretical underpinning in the $x_F$ range of
the present measurement, it nonetheless gives a good representation of
the data. The parameterization of the $p_T$ dependence is also often
seen in the literature, except that we have found it necessary to
introduce a linear term in $|x_F|$ into the factor dividing $p_T^2$
since the distributions tend to flatten with decreasing $x_F$. This is
the well-known ``sea-gull'' effect~\cite{seagull1} first
observed~\cite{seagull2} in bubble chamber experiments.  The fitted
curves are shown as dark solid lines in Fig.~\ref{fig:DoubleCS} and
the fit parameters together with the fit $\chi^2$s are summarized in
Table~\ref{tab:Fit_Par}. 
The functions are in agreement with the data at the level of 5\%
or better in the high statistics bins and otherwise compatible with
the data within statistical errors.

The reported values for the parameter $n$
are for the most part considerably larger than either those expected
by the counting rules given in \cite{brodsky} or the measurements
summarized in the same paper. However, as noted above, the model
of~\cite{brodsky} applies only for $x_F$ values outside the measured
$x_F$ range. Both Pythia and EPOS indicate that $n$ is a strong
function of $x_F$ with $n$ close to the numbers reported in
Table~\ref{tab:Fit_Par} for $|x_F| \lesssim 0.1$ but decreasing to values
similar to those given in~\cite{brodsky} for $|x_F| \approx 0.5$.  The
fitted functions have been used to calculate the values of the doubly
differential cross section in the unmeasured bins of the grid. These
values are presented in Table~\ref{tab:CS_K0s}, \ref{tab:CS_Lam} and
\ref{tab:CS_Lam_bar} (marked by asterisks). 

The results of PYTHIA and EPOS are indicated in
Fig.~\ref{fig:DoubleCS} by light solid lines and dotted lines,
respectively. The PYTHIA results are for proton-proton collisions at
$\sqrt{s} = \unit[41.6]{GeV}$ (with default settings) and therefore
the total calculated cross sections do not correspond to the measured
pA cross sections. Thus, to facilitate the comparison of shapes, the
normalizations are arbitrarily adjusted such that the PYTHIA results
agree with the data in the highest $x_F$ and lowest $p_T^2$ bin of
each plot separately.  In contrast, EPOS provides the cross section
relative to the total inelastic cross section for each target. The
inelastic cross sections are taken from~\cite{bru07}.  

As expected, the EPOS calculations generally give a better description of the data
than the (arbitrarily normalized) PYTHIA curves although PYTHIA is
remarkably good at describing the \ks\ data for the lighter target
materials.  Since the PYTHIA calculations are for proton-proton interactions,
they can be expected to give a progressively poorer description of the
data with increasing A, at least in part due to the Cronin
effect~\cite{cro77}: the flattening of the $p_T$
distribution with increasing atomic mass number.  In general, the EPOS
curves give a quite satisfactory description of the data (to better
than $\approx 20\%$ for most of the measured range) although there
is a pronounced tendency to overestimate the cross section at
low-$p_T$, particularly for the lighter targets.

\begin{figure}
  \includegraphics[clip,width=0.4\textwidth, bb=0 0 580 540]
  {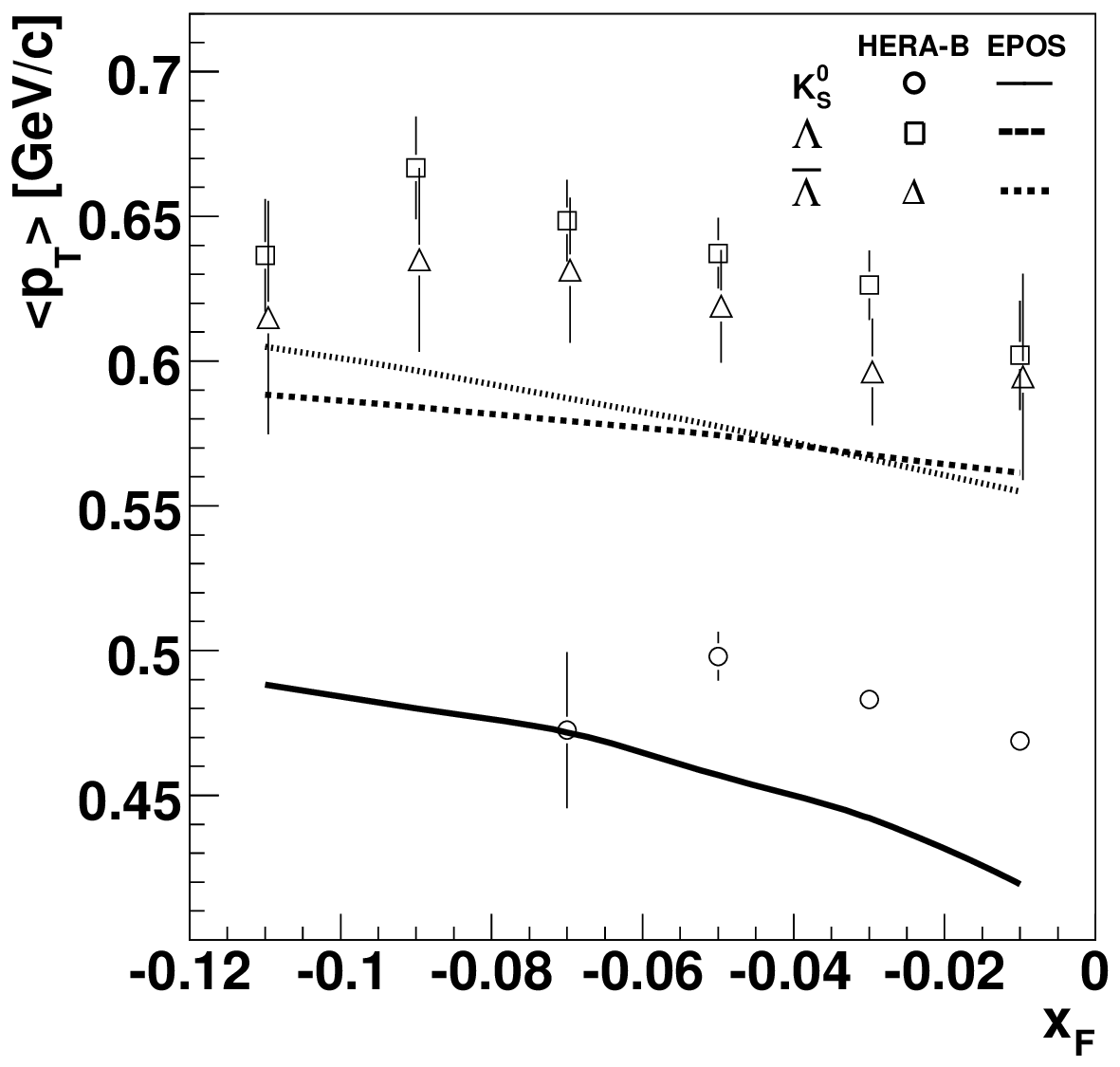}
  \caption{ The average transverse momentum, $\langle p_T \rangle$, 
    of \ks, \lam, and \lamb\ as a function of $x_F$ from the tungsten
    target sample (points) together with the EPOS data (lines).
    Error bars are statistical only.}
  \label{fig:Average_pT}
\end{figure}

\begin{figure}[h]
  \includegraphics[clip,width=0.4\textwidth, bb=0 0 560 540]
  {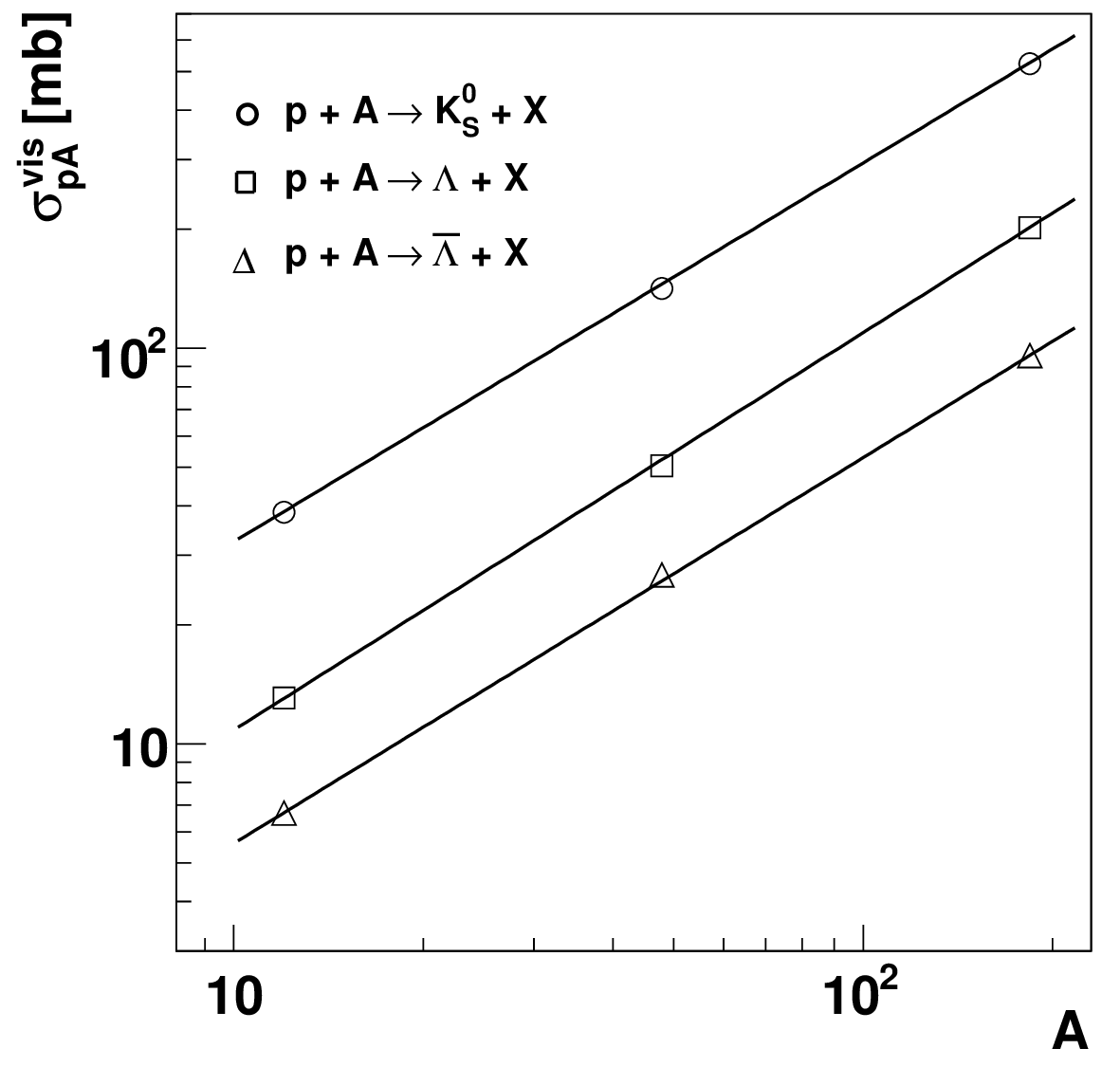}
  \caption{
    Atomic mass number dependences of the $V^{0}$ integrated inclusive
    cross sections $\sigma_{\it pA}^{\it vis}$. The solid lines
    represent fits to the parameterization
    (Eq.~\ref{eq:A_depend_vis_form}). The uncertainties include both statistical
    and systematic contributions.}
  \label{fig:A_dep_CS_vis}
\end{figure}

The average transverse momentum in a specific ($x_{F_{i}}$) slice
can be calculated using the formula:

\begin{equation}
  \label{eq:pT_Ev_fit_func}
  \langle p_{T_{i}} \rangle = \frac{ \sum_{j=1}^{n} \langle p_{T} \rangle_{i,j} \cdot \sigma_{i,j} } { \sum_{j=1}^{n}\sigma_{i,j} },
\end{equation}
where the average $p_{T}$ in the $(i,j)$'th bin, $\langle p_{T_{i,j}}
\rangle$, is calculated from the parameterization (Eq. \ref{eq:Comb_xF_pT2_fit_func}), $\sigma_{i,j}$ is
the value of the cross section in the same bin, and $\it {n}$ is the
number of ($p_{T}^{2}$) bins.  This quantity is plotted in
Fig.~\ref{fig:Average_pT} as a function of $x_{F}$ for \ks , \lam\ ,
and \lamb\ for the tungsten target sample together with the
corresponding EPOS predictions.  The EPOS predictions show the same
trend of increasing $\langle p_T \rangle$ with decreasing $x_F$ as the
data and also the same ordering with $\langle p_T \rangle$: $\langle p_T \rangle$ of \lam\ 
slightly higher than the $\langle p_T \rangle$ of \lamb\  which is higher than the
$\langle p_T \rangle$ of \ks, although the averages are slightly underestimated. The average
$p_T$ from carbon and titanium samples behave similarly (not shown).

\subsection{Integrated cross section and atomic mass number dependence}

\begin{figure*}
  \begin{center}
    \subfigure[\ks]{
      \includegraphics[clip,width=0.3133\textwidth]
      {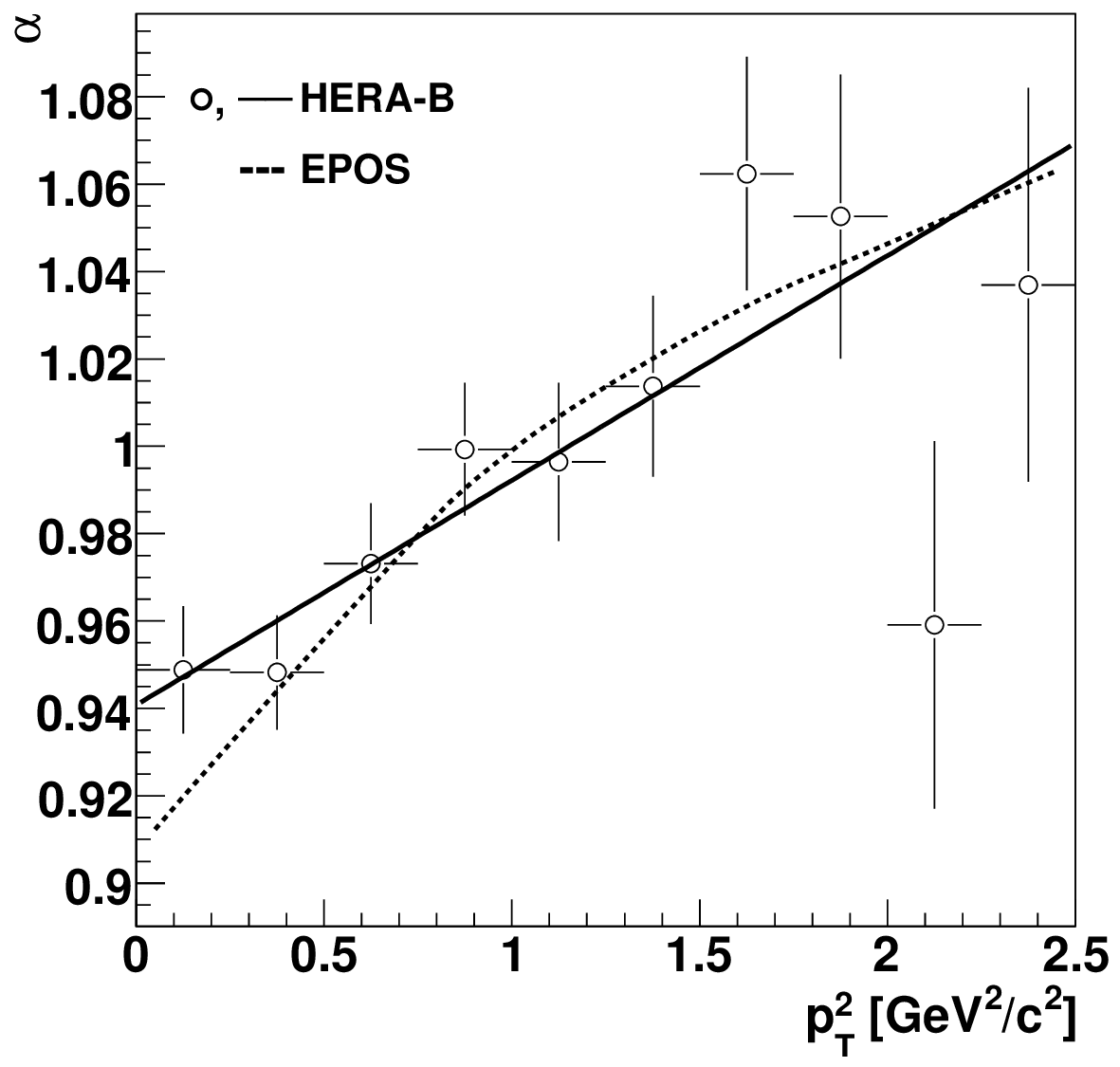}}     
    \subfigure[\lam]{
      \includegraphics[clip,width=0.3133\textwidth]
      {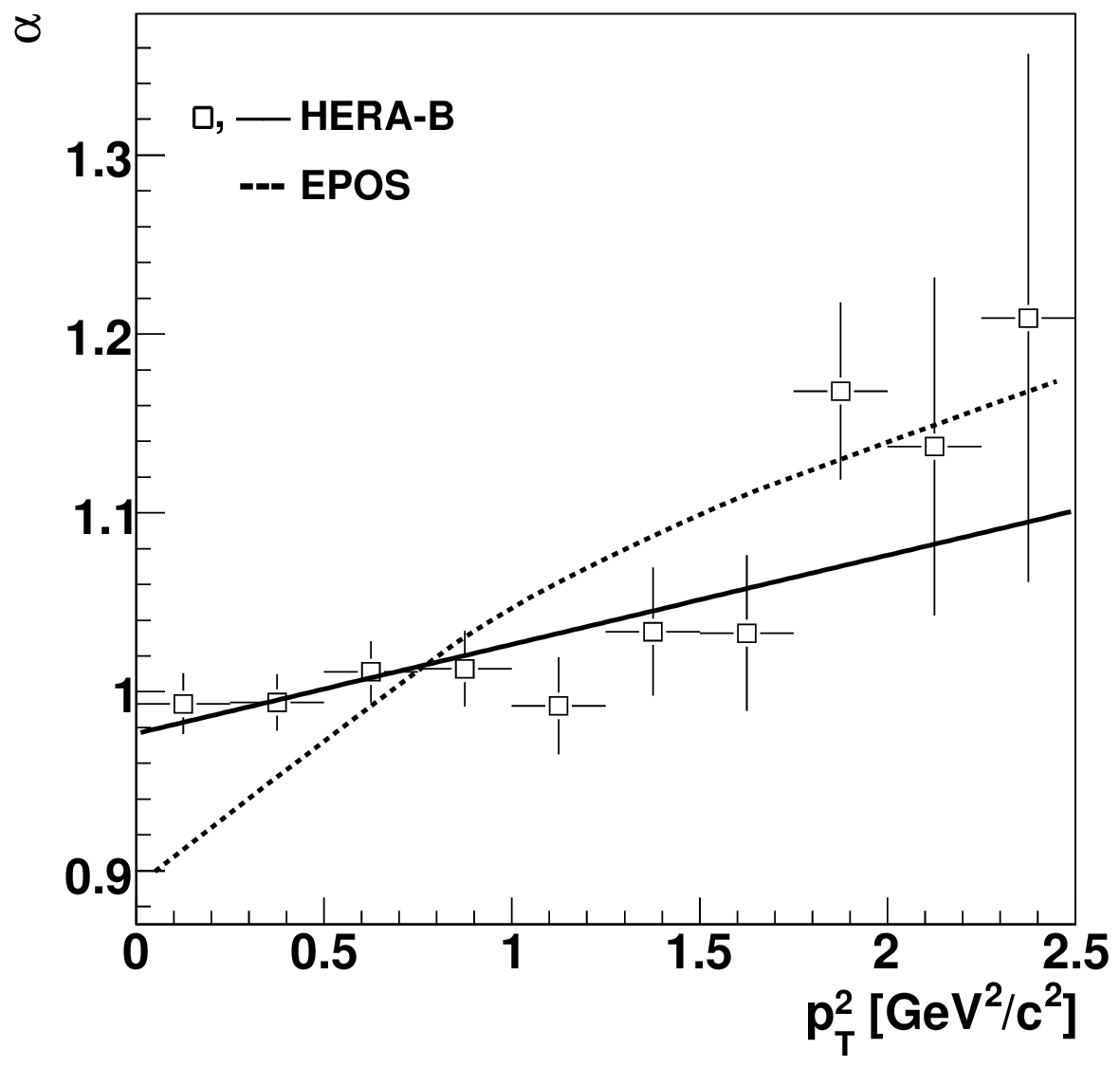}}
    \subfigure[\lamb]{
      \includegraphics[clip,width=0.3133\textwidth]
      {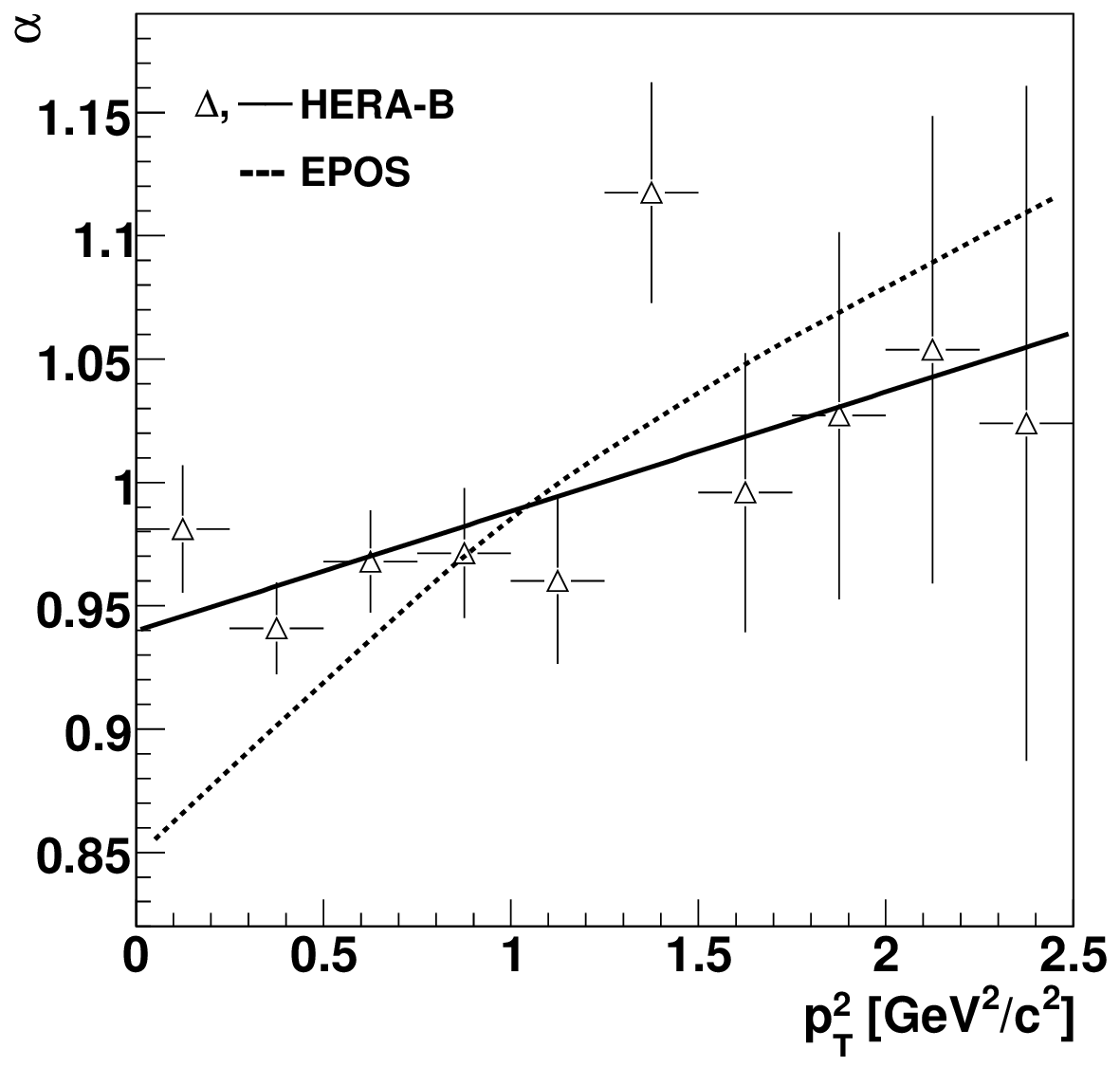}}
    \subfigure[\ks]{
      \includegraphics[clip,width=0.3133\textwidth]
      {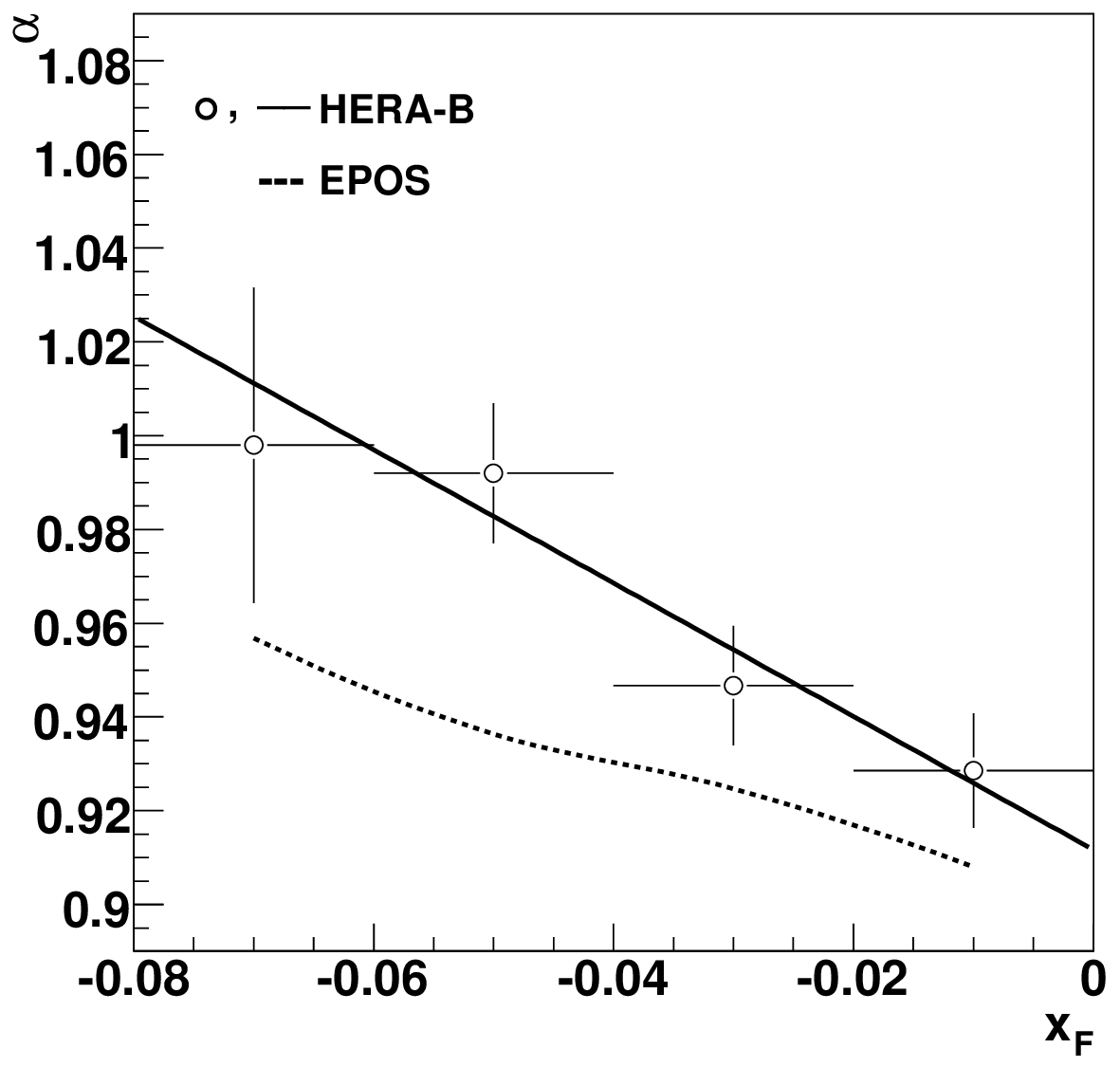}}
    \subfigure[\lam]{
      \includegraphics[clip,width=0.3133\textwidth]
      {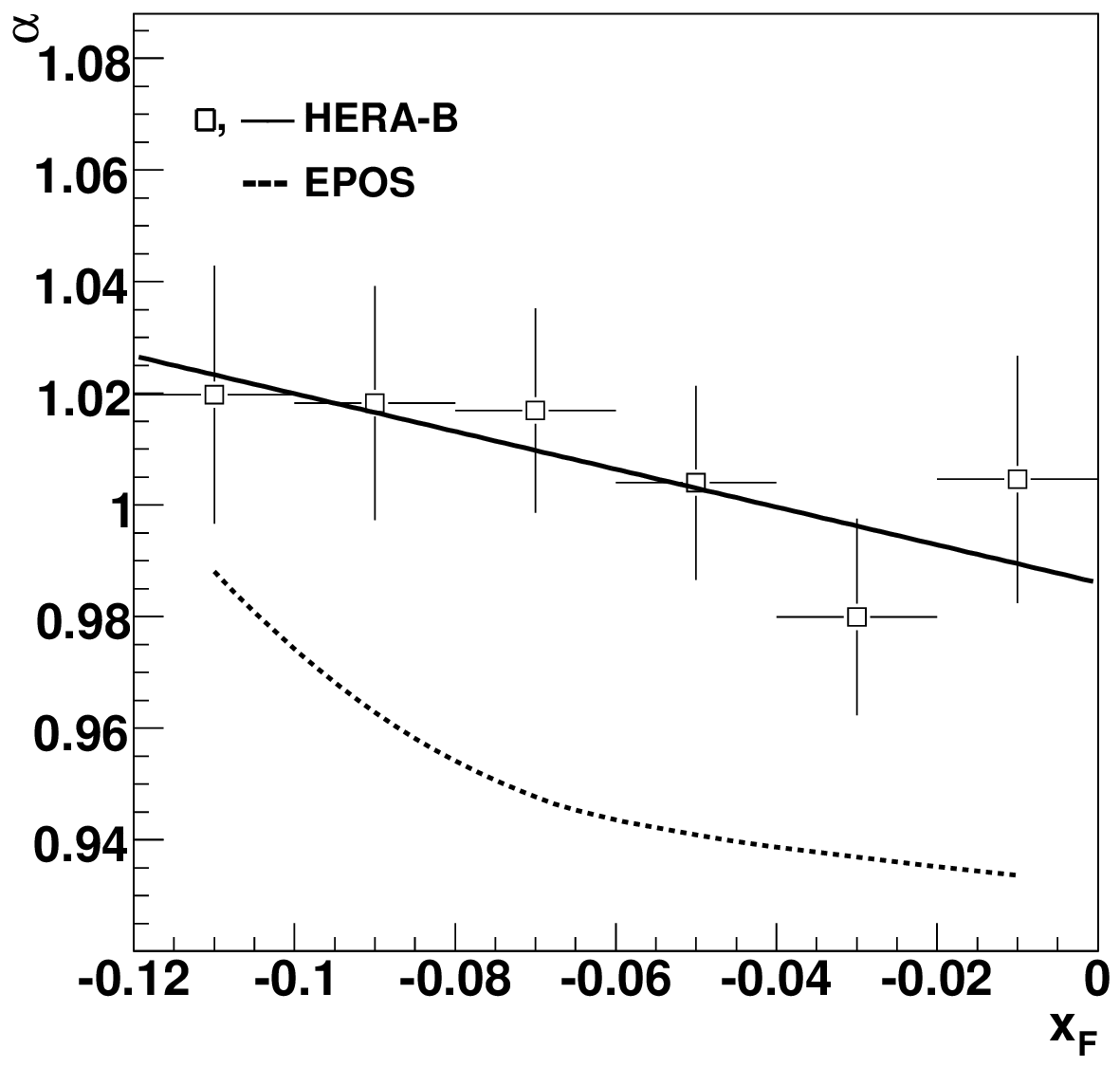}}
    \subfigure[\lamb]{
      \includegraphics[clip,width=0.3133\textwidth]
      {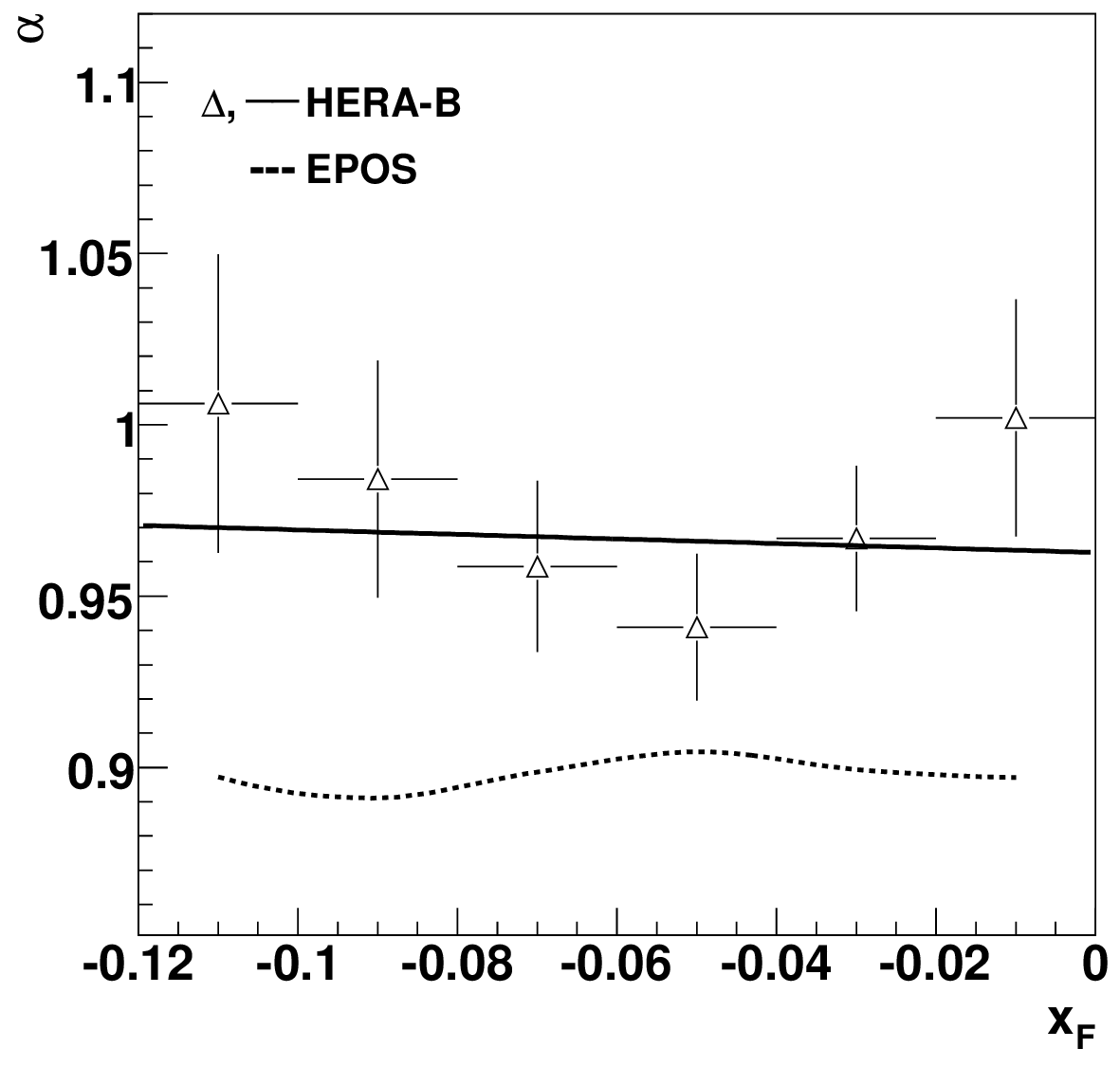}}
  \end{center}
\caption{
  The dependence of $\alpha$ for \ks, \lam\ and \lamb\ production on
  $p_T^2$ (top plots) and $x_F$ (bottom plots). The points show the
  measured values and the solid lines are the results of straight-line
  fits to the data. EPOS calculations are shown as dotted lines.  The
  uncertainties include both statistical and systematic
  contributions.}
\label{fig:Alpha_vs_xF_pT}
\end{figure*}

\begin{table}
  \begin{center}
  \caption
  {The integrated inclusive differential production cross sections
    $\sigma_{\it pA}^{\it vis}$ in the acceptance of the measurement.
    The uncertainties are statistical. The acceptance boundaries of
    the measurement in $x_F$ and $p_T^2$ are given in the 3rd and 4th
    columns, respectively.  }
  \label{tab:Tot_Int_CS}
  \begin{tabular}{@{\hspace{1mm}}l | @{\hspace{2mm}}c | @{\hspace{1mm}}c | @{\hspace{1mm}}c }
    \hline\noalign{\smallskip}
    & $x_F$ interval & $p_T^2$ range, [GeV$^2/c^2$]& $\sigma^{\it vis}_{\it pA}$, [mb] \\
    \hline\noalign{\smallskip}
    \multicolumn{4}{c}{\hspace{10mm}$p + A \to K_{S}^{0} + X$}\\
    \hline\noalign{\smallskip}
	$C$  &                  &                & $38.5\pm0.4$ \\
	$Ti$ & -$0.08$ -- $0.0$ & $0.0$ -- $2.5$ & $141.8\pm1.9$ \\
	$W$  & $ $              & $ $            & $523.9\pm5.4$  \\
    \hline\noalign{\smallskip}
    \multicolumn{4}{c}{\hspace{10mm}$p + A \to \Lambda + X$}\\
    \hline\noalign{\smallskip}
	$C$  & $ $              & $ $            & $13.1\pm0.2$ \\
	$Ti$ & -$0.12$ -- $0.0$ & $0.0$ -- $2.5$ & $50.5\pm0.7$ \\
	$W$  & $ $              & $ $            & $201.7\pm2.1$ \\
    \hline\noalign{\smallskip}
    \multicolumn{4}{c}{\hspace{10mm}$p + A \to \bar{\Lambda} + X$}\\
    \hline\noalign{\smallskip}
	$C$  & $ $              & $ $            & $6.7\pm0.2$ \\
	$Ti$ & -$0.12$ -- $0.0$ & $0.0$ -- $2.5$ & $26.7\pm0.6$ \\
	$W$  & $ $              & $ $            & $95.7\pm1.9$ \\
    \hline\noalign{\smallskip}
  \end{tabular}
 \end{center}
\end{table}

The inclusive production cross section in the visible region is
computed by summing the differential cross sections over all bins.
The results, $\sigma_{\it pA}^{\it vis}$, are listed in
Table~\ref{tab:Tot_Int_CS}.  According to the fitted functional forms,
the measured cross sections correspond to more than 98\% of the total
cross section in the visible $x_F$ interval for all targets and all
$V^0$ particles.

\begin{table}
  \caption{
    The integrated $V^0$ production cross sections per nucleon 
    $\sigma_{\it pN}^{\it vis}$ in millibarns in the visible region and the values of $\alpha$ 
    from the fit of the cross sections per nucleus to 
    Eq.~\ref{eq:A_depend_vis_form}. The uncertainties include both statistical and 
    systematic contributions. The results of fits to the data points in 
    Fig.~\ref{fig:Alpha_vs_xF_pT} are also given.}
  \label{tab:Alpha}

  \begin{center}

    \begin{tabular}{l | c | c | c} 
      \hline\noalign{\smallskip}
                                        & $K_{S}^0$       &  $\Lambda$      & $\bar{\Lambda}$\\
      \hline\noalign{\smallskip}
      $\sigma_{\it pN}^{\it vis}${\tiny [mb]} & $3.56\pm0.33$   & $1.07\pm0.11$   & $0.594\pm0.080$\\
      \hline\noalign{\smallskip}
      $\sigma_{\it pN}^{\it tot}${\tiny [mb]} & $10.33 \pm 0.90$& $6.13\pm0.61$   & $1.68\pm0.21$  \\
      \hline\noalign{\smallskip}
      $\alpha^{\it vis}$                & $0.957\pm0.013$ & $1.004\pm0.016$ & $0.975\pm0.021$\\
      \hline\noalign{\smallskip}
      $\chi^2$                          & $0.4$           & $0.9$           & $0.5$          \\
      \hline\noalign{\smallskip}
      \hline\noalign{\smallskip}
      \multicolumn{4}{c}{Fits of Figs.~\ref{fig:Alpha_vs_xF_pT}a,b,c to $\alpha = \alpha_0^{p_T} + \alpha_1^{p_T} \cdot 
p_T$}\\
      \hline\noalign{\smallskip}
      $\chi^2/_{\mbox{\tiny DOF}}$                   & $8.4/8$         &    $7.8/8$      &   $10./8$      \\
      \hline\noalign{\smallskip}
      $\alpha_0^{p_T}$                  & $0.941\pm0.011$ & $0.975\pm0.015$ & $0.938\pm0.018$ \\
      \hline\noalign{\smallskip}
      $\alpha_1^{p_T}$                  & $0.052\pm0.005$  & $0.052\pm0.007$  & $0.052\pm0.011$  \\
      \hline\noalign{\smallskip}
      \hline\noalign{\smallskip}
      \multicolumn{4}{c}{Fits of Figs.~\ref{fig:Alpha_vs_xF_pT}d,e,f to $\alpha = \alpha_0^{x_F} + \alpha_1^{x_F} \cdot 
x_F$}\\
      \hline\noalign{\smallskip}
      $\chi^2/_{\mbox{\tiny DOF}}$                   & $0.5/2$         &    $1.5/4$      &   $3.6/4$       \\
      \hline\noalign{\smallskip}
      $\alpha_0^{x_F}$                  & $0.911\pm0.021$ & $0.986\pm0.017$ & $0.962\pm0.025$ \\
      \hline\noalign{\smallskip}
      $\alpha_1^{x_F}$                  & $-1.43\pm0.54$  & $-0.346\pm 0.007$ & $-0.072\pm 0.011$\\
      \hline\noalign{\smallskip}
     
    \end{tabular}

  \end{center}

\end{table}

The dependence of the measured cross sections $\sigma_{\it pA}^{\it
  vis}$ on the atomic mass of the target material (A) can
be described by a power-law:
\begin{equation}
  \label{eq:A_depend_vis_form}
  \sigma_{\it pA}^{\it vis} \propto \mathrm{A}^{\alpha^{\it vis}},
\end{equation}

where, in this case, $\alpha^{\it vis}$ characterizes the average
atomic mass number dependence of the visible cross section.  The
systematic uncertainties on the individual cross section measurements
are highly correlated
between the target materials, therefore the least-squares likelihood
function used to extract $\sigma_{\it pA}$ and $\alpha$ uses the full
error matrix of the measurements.  The visible cross sections,
together with the fitted curves are shown in
Fig.~\ref{fig:A_dep_CS_vis}.  The fit results and $\chi^2$s are given
in Table \ref{tab:Alpha}.

The dependences of $\alpha$ on $p_T^2$ and on $x_F$ are shown on
Fig.~\ref{fig:Alpha_vs_xF_pT}.  The solid lines are from straight-line
fits whose parameters are given in Table~\ref{tab:Alpha} and the
dashed lines are the EPOS predictions.  The Cronin effect manifests
itself as an increase of $\alpha$ with increasing $p_T$.  The EPOS curves
reproduce the $p_T$ dependence rather well except for the first $p_T$
bins where EPOS underestimates $\alpha$. Since the main contributions
to the cross sections are at low $p_T$, the EPOS predictions lie well
under the data points in the $\alpha$ vs. $x_F$ plots although the
trends with $x_F$ is the same within errors.

The total cross sections (also given in Table~\ref{tab:Alpha}) are
found be dividing the visible cross sections by the fraction of the
total cross section in the visible region. This fraction was estimated
using an average of EPOS results for the fractions of all $V^0$s
produced in proton-proton and proton-neutron interactions in the
measured $x_F$ interval (34.7\%, 17.5\% and 35.4\% for \ks, \lam\ and
\lamb, respectively). The alternative of separately correcting each
proton-nucleus cross section before extrapolation to $A=1$ was
rejected since it relies more heavily on the Monte Carlo.

\subsection{Particle ratios}

\begin{figure}
  \begin{center}
    \subfigure[$\bar{\Lambda} / \Lambda$ vs. $x_F$]
    { \includegraphics[clip,width=0.4\textwidth]
      {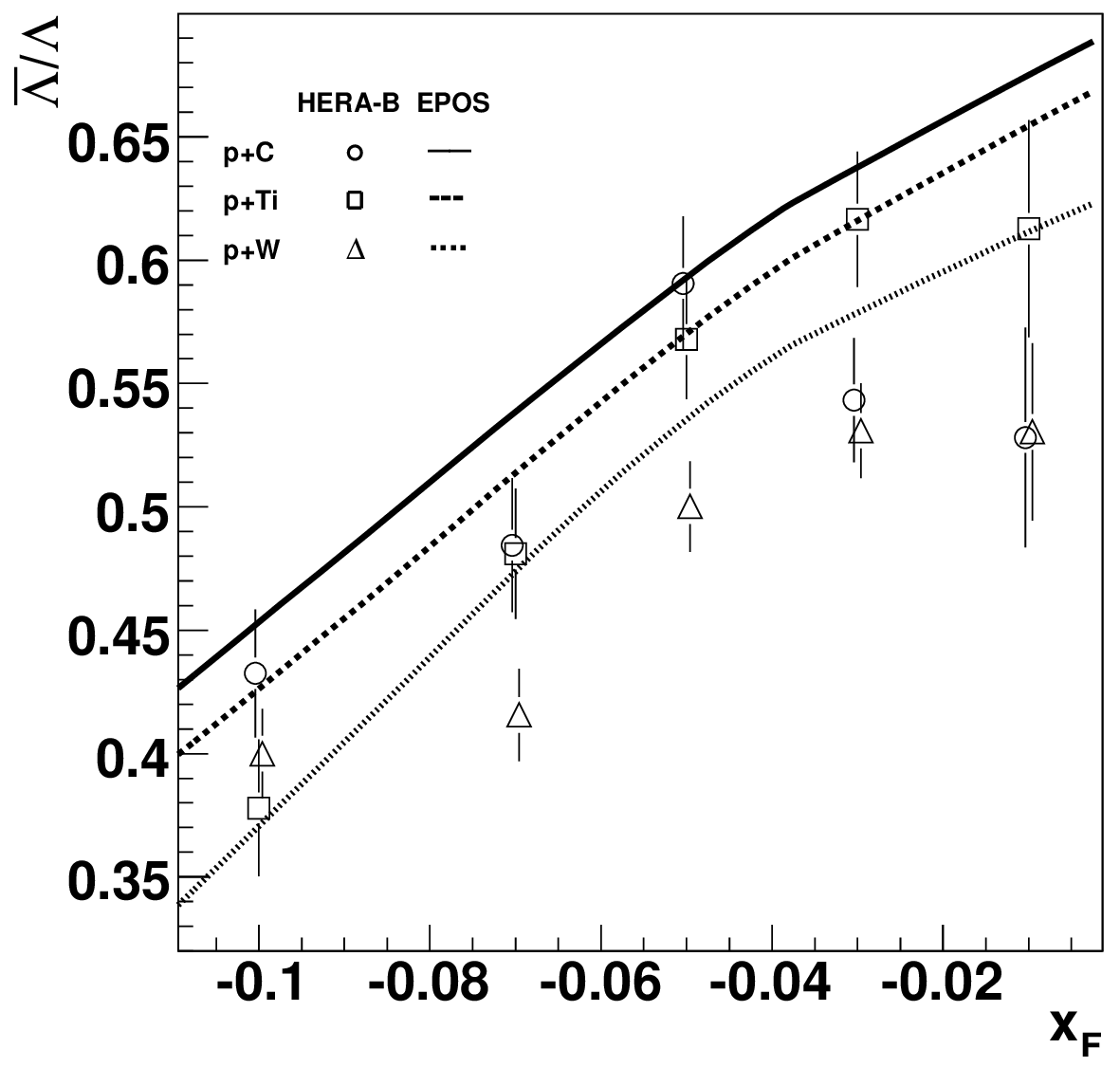}} 
    \subfigure[$\bar{\Lambda} / \Lambda$ vs. $p_T$]
    { \includegraphics[clip,width=0.4\textwidth]
      {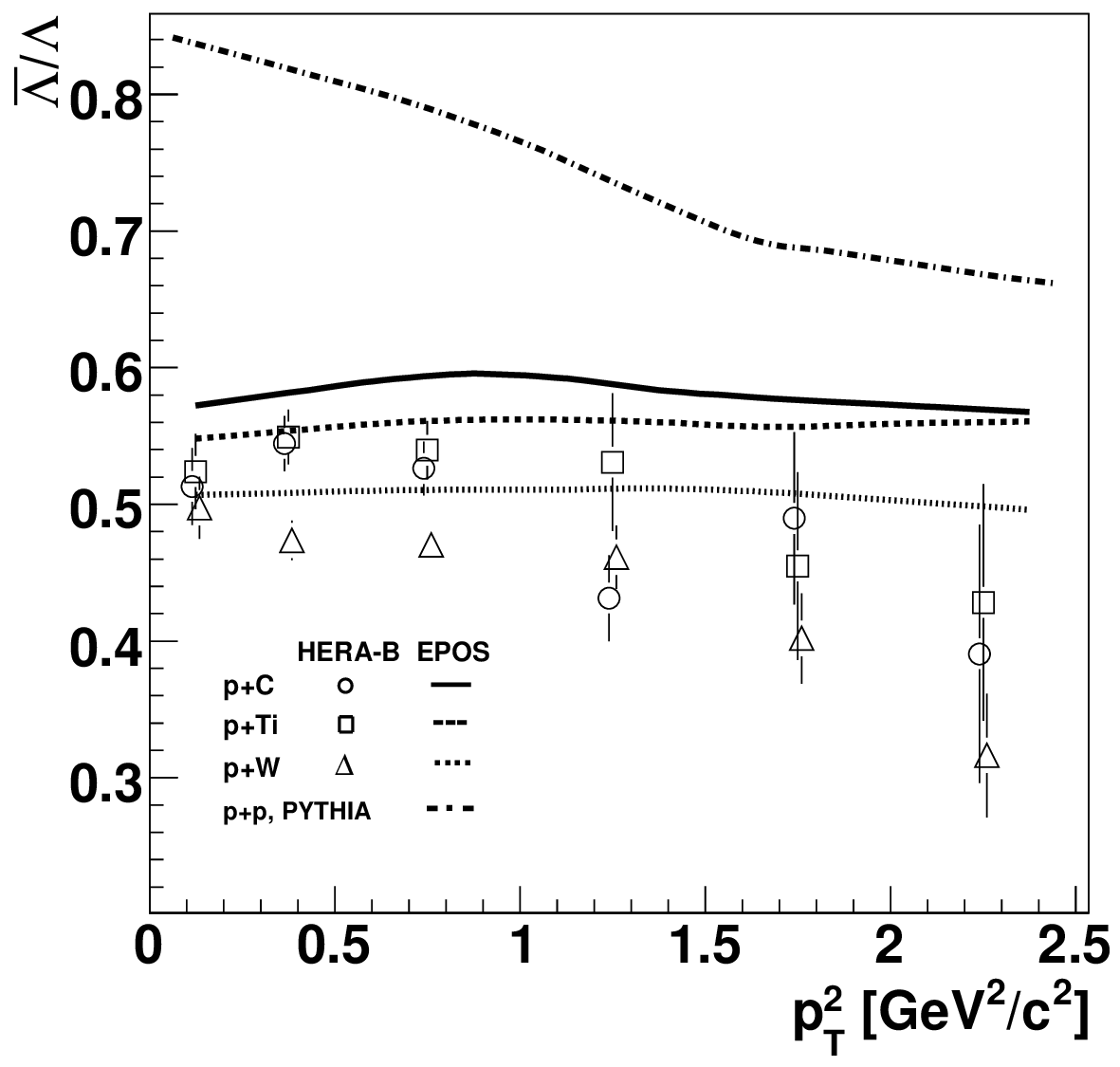}} 
  \end{center}
  \caption{The ratios of $\bar{\Lambda} / \Lambda$: a) vs. $x_F$ and
    b) vs. $p_T^2$. The points show the measured values and
    the various lines indicate the predictions of PYTHIA (part b only) and
    EPOS. The PYTHIA prediction corresponding to part (a) is well above
    the upper plot boundary (see text). The error bars show
    statistical uncertainties only. }
  \label{fig:Ratio_vs_pT_and_xF} 
\end{figure} 

\begin{figure}
  \includegraphics[clip,width=0.49\textwidth, bb=0 0 580 540]
  {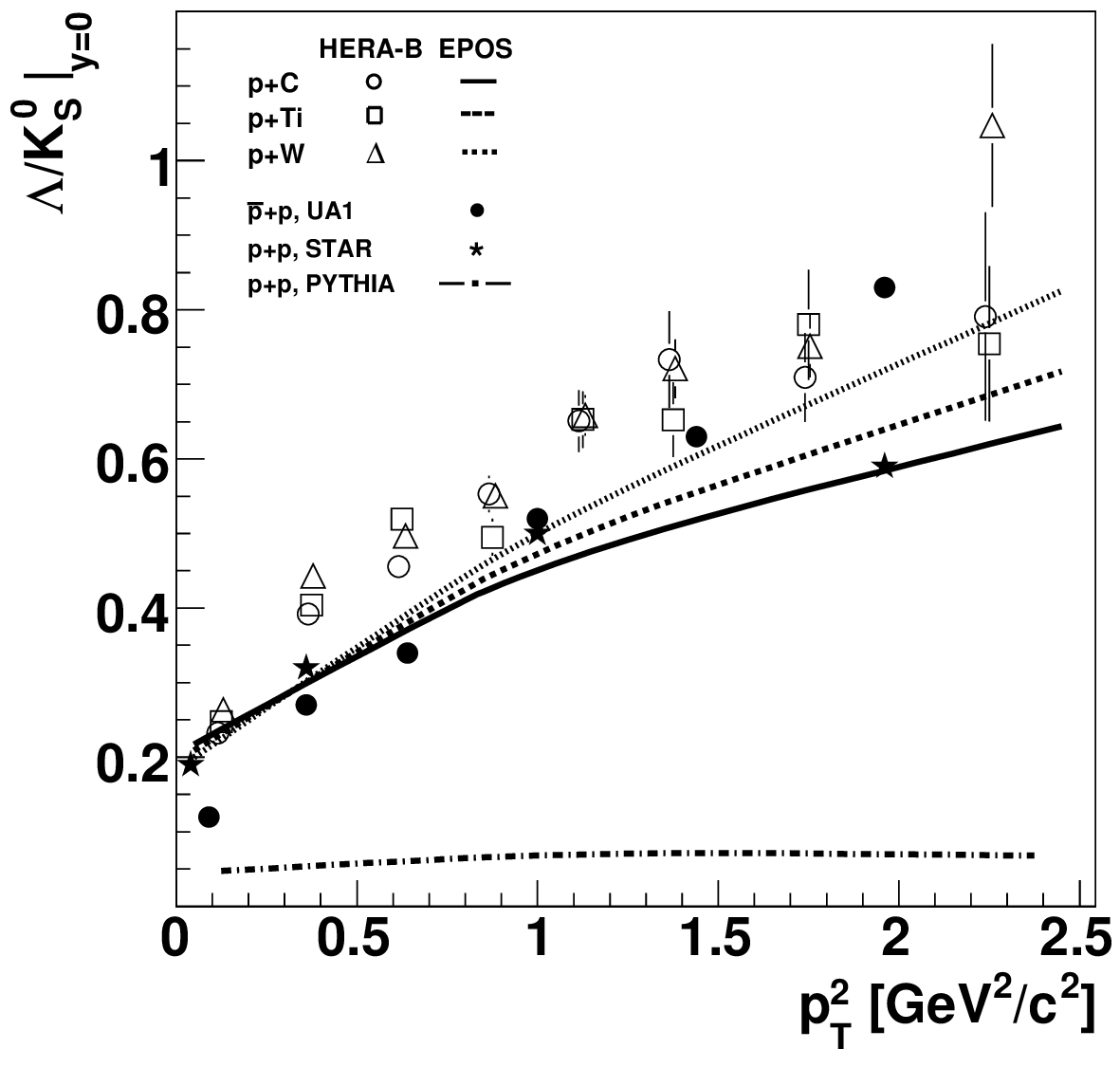}
  \caption{The ratios of $\Lambda / K_{S}^{0}$ vs. $p_T^2$ at $y \sim 0$. 
    The open points show the measured values and the solid points show
    the results from STAR and UA1 collaborations. The various
    lines indicate the predictions of PYTHIA and EPOS. The
    error bars include only statistical contributions.}
  \label{fig:Ratio_l0_k0s}
\end{figure}

The ratio of the \lamb\ cross section to that of the \lam\ is plotted in
Fig.~\ref{fig:Ratio_vs_pT_and_xF} as functions of $x_F$ and $p_T^2$
for the three targets. For Fig.~\ref{fig:Ratio_vs_pT_and_xF}a, the
data have been summed over the full measured $p_T^2$ range, and for
Fig.~\ref{fig:Ratio_vs_pT_and_xF}b, over the full $x_F$ range. The
EPOS calculations are also shown. The PYTHIA result indicated in
Fig.~\ref{fig:Ratio_vs_pT_and_xF}b, is well above the data. The PYTHIA
result vs.  $x_F$ is well above the upper plot boundary in
Fig.~\ref{fig:Ratio_vs_pT_and_xF}a, starting at {$\approx 0.8$ at
  $x_F\approx -0.1$, and increasing smoothly to $\approx 0.92$ at
  $x_F\approx 0$.  The EPOS result is in reasonable agreement
  with the data in Fig.~\ref{fig:Ratio_vs_pT_and_xF}a, where it is
  also seen to reproduce the A-dependence fairly well, despite the
  fact that the EPOS calculation of average $\alpha$ is well below the
  data for both \lam\ and \lamb\ (see Figs.~\ref{fig:Alpha_vs_xF_pT}b
  and ~\ref{fig:Alpha_vs_xF_pT}c).  As illustrated in
  Fig.~\ref{fig:Ratio_vs_pT_and_xF}b, the EPOS curve is also in reasonable
  agreement with the data over most of the $p_T^2$ range but the data
  shows a tendency to decrease with $p_T^2$ while EPOS suggests a flat
  $p_T^2$ dependence.
  
  The ratio of \lam\ to \ks\ cross sections is shown in
  Fig.~\ref{fig:Ratio_l0_k0s} for the three target materials. The STAR
  measurements~\cite{star_pp} in $pp$ interactions at $\sqrt{s} =
  \unit[200]{GeV}$ and UA1~\cite{ua1_96} in $\overline{p}p$ interactions at
  $\sqrt{s} = \unit[630]{GeV}$ are also shown. The ratio shows no appreciable
  dependence on center-of-mass energy, atomic number or the type of
  colliding particles over the measured range.  The EPOS results agree
  well with the data at low $p_T$ but tend to underestimate the data
  at higher $p_T$.  Nonetheless, as indicated in the figure, the EPOS
  calculation lies far closer to the data than the PYTHIA result over
  the full measured range.

\subsection{Comparison with existing data}

Only two experiments~\cite{busser76, drij81} have measured $V^0$
production at a similar energy and in kinematic ranges which overlap
with the present measurement. The first of these measurements, by
B\"{u}sser et al., gives the average invariant cross section as a
function of $p_T$ of three separate measurements at $\sqrt{s} = 30.6$, $44.8$,
and $ 52.7\,\mathrm{GeV}$ (an average energy of $44\,\mathrm{GeV}$) in
proton-proton collisions and in a center-of-mass rapidity ($y$)
interval of about 2 units centered at 0 and for $p_T$ larger than
\unit[1.2]{GeV/c}(\ks) and \unit[0.8]{GeV/c} (\lam\ and \lamb).  The measurements
are shown in Fig.~\ref{fig:comp_buesser}.  The second report, by
Drijard et al.~\cite{drij81} gives invariant cross sections for \ks,
\lam, and \lamb\ over a wide range in rapidity and $p_T$ in
proton-proton collisions at $\sqrt{s}=63$\,GeV. The relevant points are also
shown in Fig.~\ref{fig:comp_buesser}.  The HERA-B measurements are
indicated in Fig.~\ref{fig:comp_buesser} by curves which are derived
from the parameterization given by (Eq.~\ref{eq:pT_Ev_fit_func}). The
fit parameters are fixed to those of the carbon target
(Table~\ref{tab:Fit_Par}) and the resulting values are extrapolated to
$A=1$ assuming the straight line fits to the $\alpha$ vs. $p_T$ points
shown in Fig.~\ref{fig:Alpha_vs_xF_pT}.

While the \ks\ cross sections of~\cite{busser76} are in rather good
agreement with the HERA-B results, the HERA-B \lam\ and \lamb\ 
measurements are somewhat higher.  B\"{u}sser et al. also extrapolate
their measurements to $p_T=0$ and report $(\frac{d \sigma}{dy})_{y=0}
= 0.43 \pm 0.05$\,mb (\lam) and $0.27 \pm 0.04$\,mb (\lamb). The
corresponding numbers for the present measurement, $ (\frac{d
  \sigma}{dy})_{y=0} = 0.77 \pm 0.05$\,mb (\lam) and ($0.47 \pm
0.04$\,mb (\lamb) are nearly a factor of two higher.  As shown in
Fig.~\ref{fig:comp_buesser}, the $y=0$ measurements of
~\cite{drij81db} are also about a factor of two higher than the
present measurement. This is, at least in part, explained by the
substantially higher center-of-mass energy of the Drijard et al.
measurements, however possible problems with the \ks\ measurements
reported in~\cite{drij81} have been noted~\cite{v0paper} elsewhere.

Finally, in Fig.~\ref{fig:stot}, we show the HERA-B results together
with previously published values of the total proton-nucleon cross
section as a function of squared CM energy ($s$).  The HERA-B results
fit with the general trend of the data.  Two notable exceptions are
the two points at $s = 2800\,\mathrm{GeV}$ and $3800\,\mathrm{GeV}$
indicated by squares (\lam )and triangles (\lamb ) from Erhan et
al.~\cite{erhan79} for \lam\ and \lamb\ production. We note however
that these points depend sensitively on the extrapolations of
B\"{u}sser et al.\cite{busser76} and that a multiplicative factor of
two is missing from the transformation given in~\cite{erhan79} of the
B\"{u}sser et al. points from $d\sigma/dy$ to
$d\sigma/d|x_F|$~\cite{erhan07}.  If our own measurements are
substituted for the B\"{u}sser et al.  extrapolations, we estimate
that the total cross section values of Erhan et al. would increase by
about a factor of two and a more satisfactory agreement among the
different measurements would result, as indicated by the recalculated
points in the figure.

\begin{figure}
  \includegraphics[clip,width=0.49\textwidth, bb=0 0 580 540]
  {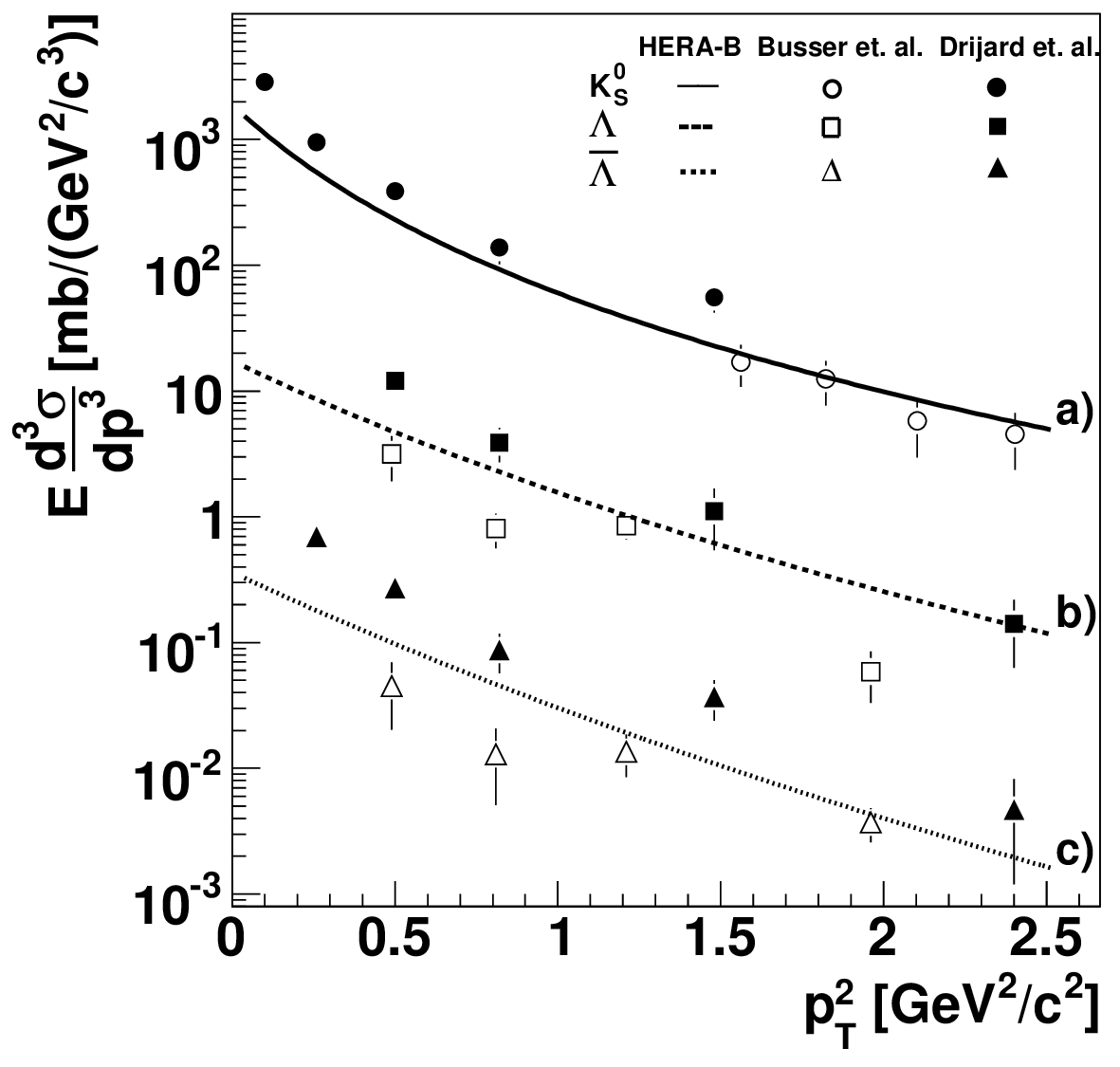}

  \caption{The average invariant cross section in the $|y| \lesssim 1$
    interval for \ks, \lam, and \lamb\ multiplied by the following
    scale factors: a)\,600, b)\,30, c)\,1.  The points are
    from~\cite{busser76} and ~\cite{drij81db}.  The curves correspond
    to parameterizations of the present measurements as explained in
    the text.
  \label{fig:comp_buesser}}
\end{figure}

\begin{figure}
  \begin{center}
    \subfigure[]{ \includegraphics[clip,width=0.45\textwidth]
      {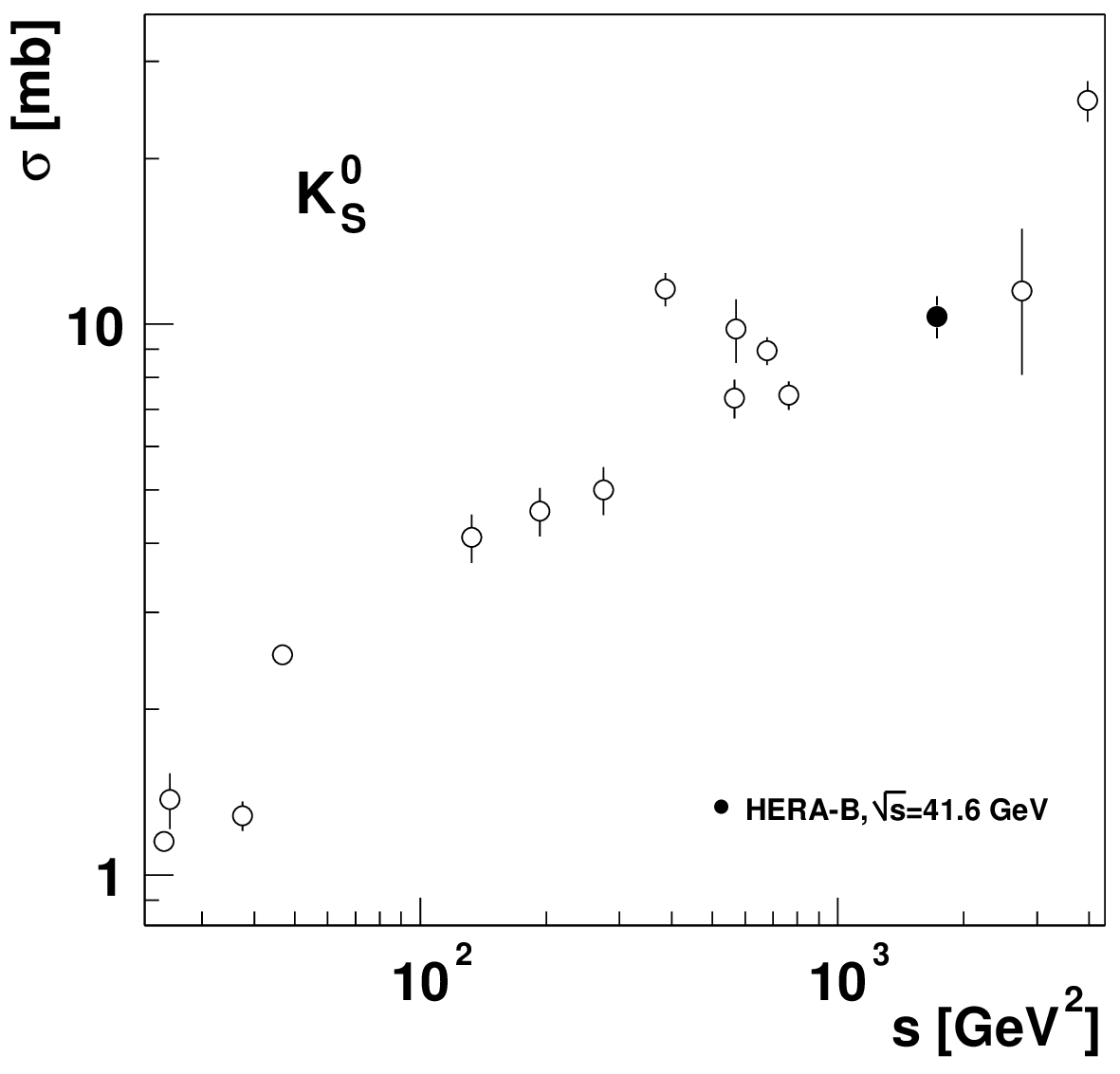}} 
    \subfigure[]{ \includegraphics[clip,width=0.45\textwidth]
      {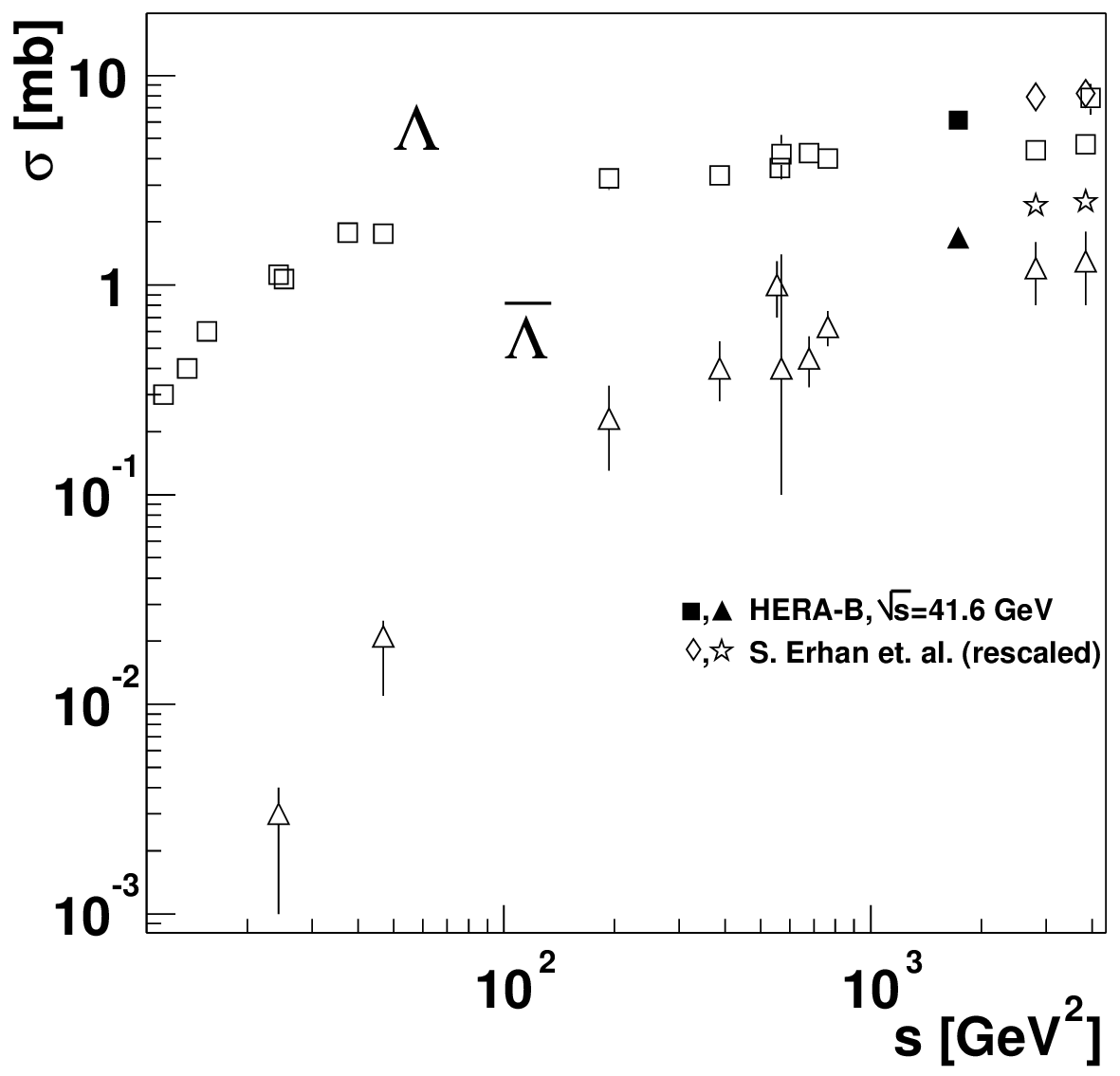}} 
  \end{center}
  \caption{A compilation of total cross section measurements from references~[6-24]
    and HERA-B for a)~\ks\ and b)~\lam\ and \lamb\ production vs.
    squared CM energy ($s$).  The recalculated data from Erhan et
    al.~\cite{erhan79} are indicated by diamonds and stars.}

  \label{fig:stot}
\end{figure}

\section{Systematic uncertainties and checks}
\label{sect:syst}

\begin{table}
  \caption{Summary of systematic uncertainties. The values are shown separately for
each particle and each target material.For the luminosity, the total and uncorrelated
errors are quoted.   \label{tab:Syst_Err}}
  \begin{center}
    \begin{tabular}{l c | c | c} 
      \hline\noalign{\smallskip}
      & \multicolumn{1}{c}{$K_{S}^{0}$} & \multicolumn{1}{c}{$\Lambda$} & \multicolumn{1}{c}{$\bar{\Lambda}$}\\
      \hline\noalign{\smallskip}
      signal counting       & $3.2\%$ & $3.3\%$  & $4.5\%$ \\
      \hline\noalign{\smallskip}
      cut variation         & $0.4\%$ & $3.9\%$  & $5.2\%$ \\
      \hline\noalign{\smallskip}
      tracking efficiency   & $3.0\%$ & $3.0\%$  & $3.0\%$ \\
      \hline\noalign{\smallskip}
      branching ratio       & $0.05\%$ & $0.5\%$ & $0.5\%$ \\
      \hline\noalign{\smallskip}
      MC model              & $3.3\% $ & $3.7\%$ & $5.7\%$ \\
      \hline\noalign{\smallskip}
      \hline\noalign{\smallskip}
      total (w/o luminosity) & $5.5\%$  & $7.0\%$ & $9.4\%$ \\ 
      \hline\noalign{\smallskip}
      \hline\noalign{\smallskip}
      & \multicolumn{1}{c}{C} & \multicolumn{1}{c}{Ti} & \multicolumn{1}{c}{W}\\
      \hline\noalign{\smallskip}
      luminosity (tot)      & $5.0\%$ & $5.2\%$ & $4.2\%$ \\
      \hline\noalign{\smallskip}
      luminosity (uncorrelated) & $3.9\%$ & $4.2\%$ & $2.9\%$ \\ 
      \hline\noalign{\smallskip}
    \end{tabular}

  \end{center}
\end{table}

The following possible sources of systematic uncertainty have been
considered:

\begin{itemize}
  
\item A bin-based method is used for estimating the number of produced
  $V^0$ candidates.  An alternative fit-based method in which the
  invariant mass distributions are fit to a double Gaussian for the
  signal and a first-order Legendre polynomial for the background
  results in changes to the cross sections of 3.2\% for \ks, 3.3\% for
  \lam\ and 4.5\% for \lamb.
  
\item From varying the most powerful cut, namely the cut on
  $\tilde{p_{T}} \cdot c\tau$, within reasonable limits, we estimate a
  systematic uncertainty of about 3.9\% for \lam, 5.2\% for \lamb\ and
  0.4\% for \ks\ mesons.
  
\item The efficiencies for reconstruction of track segments in the VDS
  and in the OTR were measured independently by exploiting $\pi^+\pi^-$
  decays~\cite{per04} of the \ks.  One of the two decay pions was
  reconstructed using RICH and ECAL information instead of either the
  OTR hits or the VDS hits and a search was made among the
  reconstructed tracks for a match.  Based on a comparison of this
  method applied to data and to Monte Carlo, a systematic uncertainty
  on track reconstruction and matching efficiency of 1.5$\%$ per track
  is estimated.
  
\item The influence of the track multiplicity on the reconstruction
  efficiency is found to give a negligible contribution to the
  systematic uncertainty.
  
\item The systematic uncertainties on the branching ratios~\cite{pdg08} are
  0.05\% for \ksto\ and 0.5\% for \lamto\ and \lambto\ decays,
  respectively.
  
\item The total systematic uncertainties due to the luminosity
  calculations~\cite{bru07} are 5.0\%, 5.2\% and 4.2\% for carbon,
  titanium and tungsten targets, respectively. The uncertainties are
  correlated between target materials with correlation coefficients
  varying between 0.90 and 0.92. For the A-dependence and pN cross
  section results, these uncertainties and their correlations are
  taken into account.
  
\item A check for a possible left-right bias in the spectrometer
  acceptance was made by deriving the visible \ks\ cross section
  with subsets of the data with opposite signs of decay asymmetry
  ($p^+_z - p^-_z)/(p^+_z + p^-_z)$, where $p^+_z$ and $p^-_z$ are the
  components of momentum along the beam direction of $\pi^+$ and
  $\pi^-$, respectively).  The maximum difference between the values
  of cross sections for the negative and positive asymmetry samples is
  $0.7\%$.
  
\item The fact that the efficiency correction was done on a grid of
  $x_F$ and $p_T$ bins considerably reduces the dependence of the
  correction on the shape of the kinematic distributions produced by
  the MC compared to separate one-dimensional corrections. The
  remaining uncertainty was studied by varying $x_F$- and
  $p_T$-dependent weighting factors applied to the MC events. The
  difference between the average efficiency computed with a weight of
  unity and a weighting map which forces FRITIOF-generated
  distributions to conform to the corrected data is taken as the
  systematic uncertainty on the MC production model. The numbers are
  given in Table~\ref{tab:Syst_Err}.
  
\item In~\cite{lampol}, we reported evidence for a positive
  polarization of \lam's relative to the normal to the \lam\ 
  production plane in the visible region. Nonetheless the acceptance
  calculations done for the present measurement assume unpolarized production of
  \lam's.  It is however also shown in~\cite{esben} that the acceptance is
  insensitive to polarization effects.
  
\item The proper lifetimes of \ks, \lam\ and \lamb\ extracted from the
  data sample are $2.65\pm0.04$\,cm, $8.70\pm0.47$\,cm and
  $8.26\pm0.68$\,cm, respectively (statistical errors only). The \ks\ 
  and \lamb\ lifetimes are within $1\,\sigma$ of the PDG
  values~\cite{pdg08} while the measured \lam\ lifetime is
  $1.7\,\sigma$ higher than the PDG value. The level of agreement is
  thus acceptable.

\end{itemize}

The systematic uncertainty estimates resulting from these
considerations are collected in Table~\ref{tab:Syst_Err}. The
systematic uncertainties on the differential cross section
measurements are quadratic sums of luminosity-dependent and $V^0$-type
dependent terms and are largely correlated over the measured range and
constant to within about 20\%. Since the uncertainties are for
the most part correlated and constant, they appear as uncertainties in
the overall scale depending only on target material and $V^0$ type and
are quoted in Tables~\ref{tab:CS_K0s},
\ref{tab:CS_Lam} and \ref{tab:CS_Lam_bar}.

\section{Summary}

We have studied the production cross sections for \ks, \lam, and
\lamb\ in the central region ($-0.12 < x_F < 0.0$) in proton
interactions on nuclear targets (carbon, titanium and tungsten) at a
center-of-mass energy of $\sqrt{s} = \mathrm{41.6\,GeV}$.  The main
results, the doubly differential cross sections are presented in
Tables~\ref{tab:CS_K0s}, \ref{tab:CS_Lam}, and \ref{tab:CS_Lam_bar}.
Several derived quantities: particle ratios, the $A$-dependence
parameter $\alpha$, and the total production cross sections are
presented and discussed.  The results are compared to PYTHIA and EPOS
calculations. For the most part, the EPOS calculations agree with the
data at the 20\% level.  PYTHIA is not designed to handle proton
nucleus interactions and, as expected, produces $p_T$ distributions
which are steeper than the data.  PYTHIA also fails to describe the
ratio of \lam\ to \lamb, and, as previously pointed out
in~\cite{star_pp}, the ratio of \lam\ to \ks.  The failure cannot be
attributed to A-dependence. The results are also compared to existing
measurements and possible reasons for some discrepancies are
discussed.

\section*{Acknowledgments}

We express our gratitude to the DESY laboratory for their strong
support in the installation of running of the experiment and the
subsequent data analysis. We are also indebted to the DESY accelerator
group for their continuous efforts to provide the best possible beam
conditions. The HERA-B experiment would not have been possible without
the enormous effort and commitment of our technical and administrative
staff. It is a pleasure to thank all of them. We thank Klaus Werner
for providing EPOS predictions and also for making the EPOS code
available to us.

\begin{table*}
 \caption{  The inclusive doubly differential cross section 
   $d^2\sigma_{pA}/dx_Fdp_T^2$ for the production of \ks\ mesons on the
   indicated targets in the given  $x_F$ and $p_T$ bins.
   The uncertainties given for each bin are statistical.
   The values marked with asterisks are extrapolated. Additional
   scale uncertainties (see Sect.~\ref{sect:syst}) are quoted 
   in the headers of each sub-table. The sums over the kinematic bins 
   in each column (row) is given in the last column (row). The
   corresponding cross section for the column (row) is the sum
   multiplied by the appropriate bin width.}

  \label{tab:CS_K0s}
   \begin{center}
    \begin{tabular}{l @{\hspace{0mm}}c | @{\hspace{0mm}}c | @{\hspace{0mm}}c | @{\hspace{0mm}}c | @{\hspace{0mm}}c | @{\hspace{0mm}}c | @{\hspace{0mm}}c }
      \hline\noalign{\smallskip}
      & \multicolumn{5}{c}{$d^2\sigma_{pA}/dx_Fdp_T^2, [\mathrm{mb/(GeV/c)^2}]$} \\
      \hline\noalign{\smallskip}
      & \multicolumn{5}{c}{$p + C \to K_{S}^{0} + X$ (scale uncertainty: $\pm 7.4\%$)} \\
      \hline\noalign{\smallskip}
      $\Delta p_T^2/\Delta x_F$ & -$0.12$ -- -$0.10$ & -$0.10$ -- -$0.08$ & -$0.08$ -- -$0.06$ & -$0.06$ -- -$0.04$ & -$0.04$ -- -$0.02$ & -$0.02$ -- $0.0$ &  sum  \\

      \hline\noalign{\smallskip}
      $0.0 - 0.25$ & $219.\pm12. ^{*}$ & $352.\pm14. ^{*}$ & $715.\pm70.$ & $829.\pm24.$ & $1354.\pm15.$ & $1973.\pm15.$ & $5443.\pm79.$ \\
$0.25 - 0.5$ & $92.8\pm3.2 ^{*}$ & $140.8\pm3.4 ^{*}$ & $210.\pm11.$ & $297.6\pm7.1$ & $425.5\pm6.6$ & $554.8\pm7.2$ & $1722.\pm17.$ \\
$0.5 - 0.75$ & $39.\pm16.$ & $55.5\pm5.8$ & $92.8\pm4.6$ & $131.8\pm4.1$ & $179.7\pm4.2$ & $207.9\pm4.4$ & $707.\pm19.$ \\
$0.75 - 1.0$ & $28.6\pm5.2$ & $31.3\pm3.1$ & $49.9\pm3.2$ & $64.9\pm2.9$ & $90.6\pm3.2$ & $99.7\pm3.2$ & $365.1\pm8.7$ \\
$1.0 - 1.25$ & $22.8\pm5.1$ & $19.8\pm2.1$ & $25.2\pm1.9$ & $38.1\pm2.2$ & $50.0\pm2.5$ & $60.9\pm2.9$ & $216.8\pm7.4$ \\
$1.25 - 1.5$ & $8.6\pm2.0$ & $10.9\pm1.5$ & $16.7\pm1.7$ & $19.3\pm1.4$ & $29.3\pm2.1$ & $31.1\pm2.1$ & $115.9\pm4.4$ \\
$1.5 - 1.75$ & $4.9\pm1.3$ & $8.9\pm1.5$ & $11.9\pm1.6$ & $15.7\pm1.6$ & $15.7\pm1.4$ & $19.3\pm1.7$ & $76.4\pm3.7$ \\
$1.75 - 2.0$ & $4.5\pm1.3$ & $7.8\pm1.6$ & $9.7\pm1.6$ & $11.0\pm1.5$ & $11.2\pm1.3$ & $9.9\pm1.1$ & $54.2\pm3.4$ \\
$2.0 - 2.25$ & $3.1\pm0.7$ & $2.8\pm0.7$ & $6.6\pm1.3$ & $6.9\pm1.0$ & $11.5\pm2.2$ & $8.7\pm1.3$ & $39.6\pm3.2$ \\
$2.25 - 2.5$ & $3.5\pm0.9$ & $4.6\pm1.5$ & $4.2\pm0.9$ & $4.2\pm0.9$ & $5.0\pm0.9$ & $5.9\pm1.1$ & $27.4\pm2.6$ \\
      \hline\noalign{\smallskip}
 sum   & $427.\pm21.$ & $635.\pm16.$ & $1143.\pm71.$ & $1418.\pm25.$ & $2173.\pm17.$ & $2971.\pm18.$ & $8767.\pm84.$ \\

%[xF] & $5794.8\pm76.1$ & $1762.3\pm42$ & $707.1\pm26.6$ & $365.1\pm19.1$ & $216.8\pm14.7$ & $115.9\pm10.8$ & $76.4\pm8.7$ & $54.2\pm7.4$ & $39.6\pm6.3$ & $27.4\pm5.2$

      \hline\noalign{\smallskip}
      & \multicolumn{5}{c}{$p + Ti \to K_{S}^{0} + X$ (scale uncertainty: $\pm 7.6\%$)} \\
      \hline\noalign{\smallskip}
      $\Delta p_T^2/\Delta x_F$ & -$0.12$ -- -$0.10$ & -$0.10$ -- -$0.08$ & -$0.08$ -- -$0.06$ & -$0.06$ -- -$0.04$ & -$0.04$ -- -$0.02$ & -$0.02$ -- $0.0$ &  sum  \\

      \hline\noalign{\smallskip}
      $0.0 - 0.25$ & $936.\pm54. ^{*}$ & $1439.\pm63. ^{*}$ & $2760.\pm330.$ & $3404.\pm120.$ & $4750.\pm61.$ & $6950.\pm52.$ & $20240\pm370.$ \\
$0.25 - 0.5$ & $386.\pm15. ^{*}$ & $562.\pm15. ^{*}$ & $745.\pm41.$ & $1124.\pm29.$ & $1523.\pm23.$ & $1989.\pm25.$ & $6328.\pm64.$ \\
$0.5 - 0.75$ & $189.5\pm5.5 ^{*}$ & $283.\pm31.$ & $348.\pm19.$ & $471.\pm16.$ & $640.\pm14.$ & $795.\pm16.$ & $2727.\pm45.$ \\
$0.75 - 1.0$ & $185.\pm38.$ & $157.\pm19.$ & $222.\pm16.$ & $249.\pm11.$ & $324.\pm11.$ & $397.\pm12.$ & $1533.\pm49.$ \\
$1.0 - 0.25$ & $57.\pm13.$ & $73.\pm11.$ & $119.\pm11.$ & $139.0\pm8.2$ & $169.3\pm8.0$ & $212.\pm10.$ & $770.\pm25.$ \\
$1.25 - 1.5$ & $37.0\pm8.7$ & $54.2\pm8.1$ & $69.4\pm6.9$ & $104.2\pm8.3$ & $112.4\pm7.4$ & $127.\pm8.$ & $504.\pm20.$ \\
$1.5 - 1.75$ & $33.8\pm8.3$ & $33.4\pm7.2$ & $48.6\pm6.6$ & $64.8\pm6.8$ & $62.3\pm5.1$ & $71.0\pm6.1$ & $314.\pm17.$ \\
$1.75 - 2.0$ & $15.8\pm5.5$ & $22.6\pm4.4$ & $33.0\pm5.2$ & $40.0\pm4.8$ & $45.0\pm4.9$ & $51.8\pm5.8$ & $208.\pm13.$ \\
$2.0 - 2.25$ & $15.8\pm5.6$ & $19.3\pm6.2$ & $22.6\pm4.4$ & $23.1\pm3.5$ & $38.2\pm5.9$ & $30.4\pm4.4$ & $149.\pm12.$ \\
$2.25 - 2.5$ & $9.8\pm4.5$ & $14.3\pm4.1$ & $12.9\pm3.4$ & $19.7\pm3.9$ & $20.8\pm3.9$ & $27.4\pm4.8$ & $105.\pm10.$ \\
      \hline\noalign{\smallskip}
   sum   & $1865.\pm70.$ & $2658.\pm76.$ & $4380.\pm340.$ & $5640.\pm130.$ & $7685.\pm69.$ & $10651.\pm62.$ & $32880.\pm390.$ \\

%[xF]  & $21511.\pm147.$ & $6468.\pm80.$ & $2728.\pm52.$ & $1533.\pm39.$ & $769.7\pm27.7$ & $504.3\pm22.5$ & $314\pm17.7$ & $208.2\pm14.4$ & $149.3\pm12.2$ & $104.8\pm10.2$ \\

      \hline\noalign{\smallskip}
      & \multicolumn{5}{c}{$p + W \to K_{S}^{0} + X$ (scale uncertainty: $\pm 6.9\%$)} \\
      \hline\noalign{\smallskip}
      $\Delta p_T^2/\Delta x_F$ & -$0.12$ -- -$0.10$ & -$0.10$ -- -$0.08$ & -$0.08$ -- -$0.06$ & -$0.06$ -- -$0.04$ & -$0.04$ -- -$0.02$ & -$0.02$ -- $0.0$ &  sum  \\

      \hline\noalign{\smallskip}
      $0.0 - 0.25$ & $3750.\pm180. ^{*}$ & $5610.\pm200. ^{*}$ & $10940.\pm980.$ & $12420.\pm340.$ & $17460.\pm180.$ & $24040.\pm150.$ & $74220.\pm1100.$ \\
$0.25 - 0.5$ & $1614.\pm50. ^{*}$ & $2284.\pm49. ^{*}$ & $2890.\pm130.$ & $4105.\pm87.$ & $5758.\pm71.$ & $7019.\pm72.$ & $23670.\pm200.$ \\
$0.5 - 0.75$ & $688.\pm205.$ & $879.\pm84.$ & $1409.\pm58.$ & $1985.\pm48.$ & $2441.\pm43.$ & $2862.\pm46.$ & $10270.\pm240.$ \\
$0.75 - 1.0$ & $478.\pm79.$ & $625.\pm47.$ & $822.\pm39.$ & $1080.\pm35.$ & $1285.\pm33.$ & $1471.\pm36.$ & $5760.\pm120.$ \\
$1.0 - 0.25$ & $246.\pm40.$ & $289.\pm26.$ & $485.\pm30.$ & $611.\pm27.$ & $729.\pm26.$ & $797.\pm27.$ & $3157.\pm73.$ \\
$1.25 - 1.5$ & $159.\pm26.$ & $243.\pm25.$ & $310.\pm24.$ & $382.\pm21.$ & $389.\pm19.$ & $458.\pm22.$ & $1940.\pm56.$ \\
$1.5 - 1.75$ & $142.\pm23.$ & $177.\pm20.$ & $225.\pm22.$ & $287.\pm21.$ & $287.\pm18.$ & $321.\pm20.$ & $1439.\pm50.$ \\
$1.75 - 2.0$ & $84.\pm16.$ & $113.\pm16.$ & $143.\pm17.$ & $181.\pm17.$ & $196.\pm17.$ & $214.\pm19.$ & $930.\pm42.$ \\
$2.0 - 2.25$ & $78.\pm15.$ & $82.\pm13.$ & $84.\pm12.$ & $103.\pm12.$ & $111.\pm12.$ & $152.\pm16.$ & $610.\pm33.$ \\
$2.25 - 2.5$ & $43.\pm11.$ & $59.\pm12.$ & $66.\pm10.$ & $87.\pm13.$ & $87.\pm11.$ & $86.\pm12.$ & $428.\pm28.$ \\
      \hline\noalign{\smallskip}
   sum   & $7280.\pm300.$ & $10360.\pm240.$ & $17370.\pm990.$ & $21240.\pm360.$ & $28740.\pm200.$ & $37420.\pm180.$ & $122420.\pm1150.$ \\

%[xF]  & $80086.\pm283.$ & $24431.\pm156.$ & $10265.\pm101.$ & $5760.\pm76.$ & $3157.\pm56.$ & $1940.\pm44.$ & $1439.\pm38.$ & $930.2\pm30.5$ & $609.6\pm24.7$ & $427.5\pm20.7$ \\

      \hline\noalign{\smallskip}
   \end{tabular}
  \end{center}
\end{table*}

\begin{table*}
 \caption
{   The inclusive doubly differential cross section 
   $d^2\sigma_{pA}/dx_Fdp_T^2$ for the production of \lam\ baryons on the
   indicated targets in the given $x_F$ and $p_T$ bins. The 
   uncertainties given for each bin are statistical.
   The values marked with asterisks are extrapolated. Additional
   scale uncertainties (see Sect.~\ref{sect:syst}) are quoted 
   in the headers of each sub-table. The sums over the kinematic bins 
   in each column (row) is given in the last column (row). The
   corresponding cross section for the column (row) is the sum
   multiplied by the appropriate bin width.}
  \label{tab:CS_Lam}
   \begin{center}
    \begin{tabular}{l @{\hspace{0mm}}c | @{\hspace{0mm}}c | @{\hspace{0mm}}c | @{\hspace{0mm}}c | @{\hspace{0mm}}c | @{\hspace{0mm}}c | @{\hspace{0mm}}c }
      \hline\noalign{\smallskip}
      & \multicolumn{5}{c}{$d^2\sigma_{pA}/dx_Fdp_T^2, {[\mathrm{mb/(GeV/c)^2}]}$} \\
      \hline\noalign{\smallskip}
      & \multicolumn{5}{c}{$p + C \to \Lambda + X$ (scale uncertainty: $\pm 8.6\%$)} \\
      \hline\noalign{\smallskip}
      $\Delta p_T^2/\Delta x_F$ & -$0.12$ -- -$0.10$ & -$0.10$ -- -$0.08$ & -$0.08$ -- -$0.06$ & -$0.06$ -- -$0.04$ & -$0.04$ -- -$0.02$ & -$0.02$ -- $0.0$ &  sum  \\

      \hline\noalign{\smallskip}
      $0.0 - 0.25$ & $123.8\pm8.5$ & $154.0\pm8.7$ & $171.7\pm8.1$ & $192.3\pm8.4$ & $261.\pm12.$ & $229.\pm22.$ & $1132.\pm30.$ \\
$0.25 - 0.5$ & $55.4\pm4.1$ & $62.6\pm4.0$ & $90.2\pm4.2$ & $99.0\pm4.2$ & $125.3\pm5.0$ & $150.9\pm7.8$ & $583.\pm12.$ \\
$0.5 - 0.75$ & $35.3\pm3.6$ & $47.0\pm3.6$ & $45.0\pm3.0$ & $55.7\pm3.3$ & $60.0\pm3.2$ & $79.3\pm4.8$ & $322.3\pm8.9$ \\
$0.75 - 1.0$ & $23.6\pm3.2$ & $27.8\pm3.2$ & $30.2\pm2.8$ & $32.1\pm2.7$ & $41.1\pm3.1$ & $47.2\pm3.8$ & $201.9\pm7.7$ \\
$1.0 - 0.25$ & $16.7\pm2.8$ & $15.3\pm2.2$ & $22.3\pm3.2$ & $31.2\pm4.0$ & $26.4\pm2.8$ & $29.2\pm3.3$ & $141.1\pm7.6$ \\
$1.25 - 1.5$ & $13.8\pm4.0$ & $10.2\pm2.2$ & $15.3\pm3.2$ & $14.5\pm2.2$ & $14.8\pm2.2$ & $16.4\pm2.2$ & $85.0\pm6.8$ \\
$1.5 - 1.75$ & $11.4\pm3.6$ & $10.2\pm2.4$ & $6.0\pm1.2$ & $9.5\pm1.7$ & $13.4\pm3.2$ & $11.4\pm2.0$ & $61.9\pm6.1$ \\
$1.75 - 2.0$ & $3.7\pm1.0$ & $5.5\pm1.8$ & $3.1\pm0.8$ & $6.1\pm1.4$ & $6.7\pm1.4$ & $5.5\pm1.2$ & $30.7\pm3.2$ \\
$2.0 - 2.25$ & $4.3\pm1.3$ & $5.8\pm3.1$ & $7.0\pm4.8$ & $6.0\pm2.0$ & $3.5\pm1.0$ & $3.8\pm1.2$ & $30.3\pm6.4$ \\
$2.25 - 2.5$ & $5.9\pm4.7$ & $2.4\pm1.0$ & $1.2\pm0.4$ & $5.9\pm3.1$ & $2.1\pm0.6$ & $5.1\pm1.8$ & $22.7\pm6.0$ \\
      \hline\noalign{\smallskip}
  sum   & $294.\pm13.$ & $341.\pm12.$ & $392.\pm12.$ & $452.\pm12.$ & $555.\pm15.$ & $578.\pm24.$ & $2612.\pm38.$ \\

      \hline\noalign{\smallskip}
      & \multicolumn{5}{c}{$p + Ti \to \Lambda + X$ (scale uncertainty: $\pm 8.7\%$)} \\
      \hline\noalign{\smallskip}
      $\Delta p_T^2/\Delta x_F$ & -$0.12$ -- -$0.10$ & -$0.10$ -- -$0.08$ & -$0.08$ -- -$0.06$ & -$0.06$ -- -$0.04$ & -$0.04$ -- -$0.02$ & -$0.02$ -- $0.0$ &  sum  \\

      \hline\noalign{\smallskip}
      $0.0 - 0.25$ & $478.\pm32.$ & $533.\pm30.$ & $654.\pm30.$ & $738.\pm30.$ & $837.\pm37.$ & $1181.\pm86.$ & $4420.\pm110.$ \\
$0.25 - 0.5$ & $243.\pm17.$ & $274.\pm15.$ & $358.\pm16.$ & $395.\pm16.$ & $440.\pm17.$ & $463.\pm23.$ & $2172.\pm43.$ \\
$0.5 - 0.75$ & $195.\pm21.$ & $166.\pm12.$ & $195.\pm12.$ & $220.\pm12.$ & $268.\pm14.$ & $274.\pm16.$ & $1318.\pm36.$ \\
$0.75 - 1.0$ & $90.\pm14.$ & $87.4\pm8.4$ & $113.\pm10.$ & $125.\pm10.$ & $151.\pm10.$ & $193.\pm16.$ & $759.\pm29.$ \\
$1.0 - 0.25$ & $53.\pm10.$ & $76.\pm11.$ & $82.\pm10.$ & $94.\pm10.$ & $92.7\pm9.2$ & $105.\pm11.$ & $502.\pm25.$ \\
$1.25 - 1.5$ & $57.\pm14.$ & $36.8\pm5.9$ & $43.9\pm6.7$ & $49.0\pm6.2$ & $72.1\pm9.5$ & $70.4\pm8.8$ & $329.\pm22.$ \\
$1.5 - 1.75$ & $27.9\pm7.5$ & $62.\pm23.$ & $27.4\pm4.8$ & $39.3\pm6.9$ & $40.6\pm7.7$ & $63.\pm13.$ & $260.\pm30.$ \\
$1.75 - 2.0$ & $11.6\pm3.8$ & $14.9\pm3.0$ & $34.\pm14.$ & $24.1\pm4.6$ & $35.7\pm9.0$ & $27.2\pm6.0$ & $147.\pm19.$ \\
$2.0 - 2.25$ & $16.8\pm8.1$ & $12.0\pm0.7 ^{*}$ & $30.\pm14.$ & $14.9\pm4.3$ & $27.1\pm7.4$ & $24.7\pm6.4$ & $126.\pm19.$ \\
$2.25 - 2.5$ & $5.9\pm2.9$ & $5.8\pm2.2$ & $17.7\pm8.0$ & $7.0\pm2.0$ & $13.7\pm5.0$ & $16.2\pm8.7$ & $66.\pm14.$ \\
      \hline\noalign{\smallskip}
  sum   & $1177.\pm48.$ & $1268.\pm45.$ & $1555.\pm45.$ & $1705.\pm40.$ & $1977.\pm48.$ & $2417.\pm94.$ & $10100.\pm140.$ \\

      \hline\noalign{\smallskip}
      & \multicolumn{5}{c}{$p + W \to \Lambda + X$ (scale uncertainty: $\pm 8.2\%$)} \\
      \hline\noalign{\smallskip}
      $\Delta p_T^2/\Delta x_F$ & -$0.12$ -- -$0.10$ & -$0.10$ -- -$0.08$ & -$0.08$ -- -$0.06$ & -$0.06$ -- -$0.04$ & -$0.04$ -- -$0.02$ & -$0.02$ -- $0.0$ &  sum  \\

      \hline\noalign{\smallskip}
      $0.0 - 0.25$ & $1910.\pm110.$ & $2075.\pm93.$ & $2584.\pm92.$ & $2900.\pm93.$ & $3452.\pm119.$ & $4080.\pm260.$ & $17000.\pm340.$ \\
$0.25 - 0.5$ & $1053.\pm58.$ & $1194.\pm53.$ & $1260.\pm44.$ & $1517.\pm47.$ & $1708.\pm50.$ & $2012.\pm80.$ & $8740.\pm140.$ \\
$0.5 - 0.75$ & $641.\pm43.$ & $736.\pm40.$ & $849.\pm40.$ & $820.\pm33.$ & $949.\pm37.$ & $1096.\pm49.$ & $5091.\pm99.$ \\
$0.75 - 1.0$ & $379.\pm35.$ & $405.\pm33.$ & $482.\pm31.$ & $585.\pm33.$ & $631.\pm32.$ & $687.\pm39.$ & $3168.\pm83.$ \\
$1.0 - 0.25$ & $276.\pm31.$ & $289.\pm27.$ & $323.\pm28.$ & $387.\pm30.$ & $411.\pm31.$ & $384.\pm27.$ & $2069.\pm71.$ \\
$1.25 - 1.5$ & $128.\pm20.$ & $190.\pm23.$ & $237.\pm27.$ & $270.\pm31.$ & $302.\pm31.$ & $271.\pm26.$ & $1398.\pm64.$ \\
$1.5 - 1.75$ & $171.\pm33.$ & $175.\pm30.$ & $199.\pm32.$ & $136.\pm16.$ & $177.\pm23.$ & $180.\pm22.$ & $1038.\pm65.$ \\
$1.75 - 2.0$ & $94.\pm25.$  & $124.\pm25.$ & $122.\pm21.$ & $136.\pm24.$ & $127.\pm21.$ & $135.\pm23.$ & $739.\pm57.$ \\
$2.0 - 2.25$ & $42.\pm11.$  & $144.\pm47.$ & $126.\pm40.$ & $131.\pm39.$ & $111.\pm30.$ & $87.\pm17. $ & $640.\pm81.$ \\
$2.25 - 2.5$ & $38.\pm17.$  & $93.\pm44.$  & $82.\pm30.$  & $70.\pm19.$  & $96.\pm24.$  & $65.\pm15.$  & $443.\pm65.$ \\
      \hline\noalign{\smallskip}
  sum   & $4730.\pm150.$ & $5430.\pm150.$ & $6270.\pm140.$ & $6950.\pm130.$ & $7960.\pm150.$ & $9000.\pm280.$ & $40330.\pm430.$ \\

      \hline\noalign{\smallskip}
   \end{tabular}
  \end{center}
\end{table*}

\begin{table*}
 \caption
   {The inclusive doubly differential cross section 
   $d^2\sigma_{pA}/dx_Fdp_T^2$ for the production of \lamb\ baryons on the
   indicated targets  in the given  $x_F$ and $p_T$ bins.
   The uncertainties given for each bin are statistical.
   The values marked with asterisks are extrapolated. Additional
   scale uncertainties (see Sect.~\ref{sect:syst}) are quoted 
   in the headers of each sub-table. The sums over the kinematic bins 
   in each column (row) is given in the last column (row). The
   corresponding cross section for the column (row) is the sum
   multiplied by the appropriate bin width.}
  \label{tab:CS_Lam_bar}
   \begin{center}
    \begin{tabular}{l @{\hspace{0mm}}c | @{\hspace{0mm}}c | @{\hspace{0mm}}c | @{\hspace{0mm}}c | @{\hspace{0mm}}c | @{\hspace{0mm}}c | @{\hspace{0mm}}c }
      \hline\noalign{\smallskip}
      & \multicolumn{5}{c}{$d^2\sigma_{pA}/dx_Fdp_T^2, {[\mathrm{mb/(GeV/c)^2}]}$} \\
      \hline\noalign{\smallskip}
      & \multicolumn{5}{c}{$p + C \to \bar{\Lambda} + X$ (scale uncertainty: $\pm 10.6\%$)} \\
      \hline\noalign{\smallskip}
      $\Delta p_T^2/\Delta x_F$ & -$0.12$ -- -$0.10$ & -$0.10$ -- -$0.08$ & -$0.08$ -- -$0.06$ & -$0.06$ -- -$0.04$ & -$0.04$ -- -$0.02$ & -$0.02$ -- $0.0$ &  sum  \\

      \hline\noalign{\smallskip}
      $0.0 - 0.25$ & $58.5\pm9.2$ & $72.6\pm8.3$ & $83.0\pm7.0$ & $109.4\pm7.5$ & $134.7\pm9.6$ & $123.\pm21.$ & $581.\pm28.$ \\
$0.25 - 0.5$ & $26.2\pm3.7$ & $31.0\pm3.2$ & $43.8\pm3.3$ & $67.2\pm3.8$ & $68.3\pm3.7$ & $81.2\pm5.6$ & $317.7\pm9.7$ \\
$0.5 - 0.75$ & $15.2\pm2.6$ & $16.6\pm2.2$ & $22.6\pm2.1$ & $32.2\pm2.6$ & $44.7\pm2.9$ & $42.3\pm3.1$ & $173.5\pm6.4$ \\
$0.75 - 1.0$ & $10.9\pm1.9$ & $8.7\pm1.4$ & $16.7\pm2.4$ & $21.9\pm2.3$ & $22.7\pm2.3$ & $21.5\pm2.3$ & $102.4\pm5.2$ \\
$1.0 - 0.25$ & $4.5\pm1.7$ & $7.4\pm1.4$ & $9.0\pm1.6$ & $14.0\pm2.1$ & $12.9\pm2.1$ & $16.1\pm1.9$ & $63.7\pm4.4$ \\
$1.25 - 1.5$ & $6.6\pm2.2$ & $3.1\pm0.8$ & $5.7\pm1.3$ & $5.5\pm1.1$ & $5.2\pm1.1$ & $7.8\pm1.3$ & $33.9\pm3.4$ \\
$1.5 - 1.75$ & $1.6\pm0.7$ & $3.2\pm1.0$ & $5.0\pm1.6$ & $5.1\pm1.3$ & $5.9\pm2.0$ & $5.1\pm1.1$ & $25.9\pm3.3$ \\
$1.75 - 2.0$ & $1.4\pm0.6$ & $2.0\pm0.8$ & $1.7\pm0.7$ & $6.3\pm2.4$ & $3.3\pm1.3$ & $4.8\pm1.7$ & $19.5\pm3.5$ \\
$2.0 - 2.25$ & $0.7\pm0.5$ & $1.1\pm0.8$ & $1.7\pm1.2$ & $4.6\pm2.1$ & $2.3\pm0.8$ & $1.3\pm0.6$ & $11.7\pm2.8$ \\
$2.25 - 2.5$ & $0.8\pm0.5$ & $2.7\pm1.2$ & $0.7\pm0.4$ & $1.0\pm0.1 ^{*}$ & $1.4\pm1.0$ & $2.3\pm1.6$ & $9.0\pm2.3$ \\
      \hline\noalign{\smallskip}
  sum   & $126.\pm11.$ & $148.\pm10.$ & $189.9\pm8.9$ & $267.\pm10.$ & $301.\pm12.$ & $305.\pm22.$ & $1338.\pm32.$ \\

      \hline\noalign{\smallskip}
      & \multicolumn{5}{c}{$p + Ti \to \bar{\Lambda} + X$ (scale uncertainty: $\pm 10.7\%$)} \\
      \hline\noalign{\smallskip}
      $\Delta p_T^2/\Delta x_F$ & -$0.12$ -- -$0.10$ & -$0.10$ -- -$0.08$ & -$0.08$ -- -$0.06$ & -$0.06$ -- -$0.04$ & -$0.04$ -- -$0.02$ & -$0.02$ -- $0.0$ &  sum  \\

      \hline\noalign{\smallskip}
      $0.0 - 0.25$ & $160.\pm32.$ & $190.\pm29.$ & $320.\pm28.$ & $445.\pm28.$ & $513.\pm35.$ & $689.\pm83.$ & $2320..\pm110.$ \\
$0.25 - 0.5$ & $82.\pm15.$ & $109.\pm12.$ & $151.\pm11.$ & $216.\pm12.$ & $278.\pm14.$ & $358.\pm22.$ & $1193.\pm37.$ \\
$0.5 - 0.75$ & $72.\pm19.$ & $81.\pm11.$  & $119.\pm12.$ & $121.0\pm8.6$ & $175.\pm12.$ & $182.\pm14.$ & $750.\pm32.$ \\
$0.75 - 1.0$ & $25.6\pm7.2$ & $39.2\pm7.0$ & $51.1\pm6.3$ & $73.7\pm7.7$ & $98.1\pm9.6$ & $83.0\pm7.9$ & $371.\pm19.$ \\
$1.0 - 0.25$ & $56.\pm31.$ & $23.8\pm4.6$ & $39.5\pm6.6$ & $42.7\pm6.2$ & $45.1\pm5.3$ & $68.\pm10.$ & $274. \pm34.$ \\
$1.25 - 1.5$ & $12.5\pm4.6$ & $32.\pm11.$ & $22.3\pm4.6$ & $28.9\pm5.1$ & $30.2\pm5.9$ & $41.6\pm7.3$ & $167.\pm16.$ \\
$1.5 - 1.75$ & $14.6\pm6.9$ & $7.7\pm2.0$ & $18.1\pm5.8$ & $14.8\pm3.4$ & $24.8\pm5.2$ & $21.9\pm5.0$ & $102.\pm12.$ \\
$1.75 - 2.0$ & $2.1\pm1.3$ & $6.3\pm0.5 ^{*}$ & $16.5\pm6.5$ & $11.2\pm3.6$ & $37.\pm18.$ & $10.6\pm3.0$ & $84.\pm20.$ \\
$2.0 - 2.25$ & $3.0\pm1.6$ & $4.1\pm0.4 ^{*}$ & $8.4\pm4.1$ & $11.3\pm3.8$ & $8.7\pm3.5$ & $21.4\pm9.9$ & $57.\pm12.$ \\
$2.25 - 2.5$ & $2.2\pm0.3 ^{*}$ & $1.9\pm0.9$ & $2.4\pm0.8$ & $4.8\pm2.1$ & $8.6\pm3.8$ & $6.2\pm3.3$ & $26.1\pm5.6$ \\
      \hline\noalign{\smallskip}
  sum  & $431.\pm52.$ & $494.\pm36.$ & $748.\pm35.$ & $968.\pm35.$ & $1219.\pm45.$ & $1481.\pm89.$ & $5340.\pm130.$ \\

      \hline\noalign{\smallskip}
      & \multicolumn{5}{c}{$p + W \to \bar{\Lambda} + X$ (scale uncertainty: $\pm 10.3\%$)} \\
      \hline\noalign{\smallskip}
      $\Delta p_T^2/\Delta x_F$ & -$0.12$ -- -$0.10$ & -$0.10$ -- -$0.08$ & -$0.08$ -- -$0.06$ & -$0.06$ -- -$0.04$ & -$0.04$ -- -$0.02$ & -$0.02$ -- $0.0$ &  sum  \\

      \hline\noalign{\smallskip}
      $0.0 - 0.25$ & $850.\pm100.$ & $876.\pm87.$ & $1093.\pm84.$ & $1534.\pm89.$ & $1950.\pm110.$ & $2160.\pm270.$ & $8460.\pm350.$ \\
$0.25 - 0.5$ & $426.\pm48.$ & $425.\pm37.$ & $546.\pm37.$ & $702.\pm35.$ & $920.\pm40.$ & $1122.\pm61.$ & $4140.\pm110.$ \\
$0.5 - 0.75$ & $245.\pm32.$ & $327.\pm30.$ & $331.\pm25.$ & $427.\pm25.$ & $523.\pm29.$ & $593.\pm34.$ & $2446.\pm72.$ \\
$0.75 - 1.0$ & $110.\pm19.$ & $145.\pm18.$ & $222.\pm23.$ & $318.\pm24.$ & $305.\pm21.$ & $338.\pm26.$ & $1438.\pm54.$ \\
$1.0 - 0.25$ & $74.\pm16.$  & $135.\pm20.$ & $121.\pm14.$ & $158.\pm18.$ & $200.\pm19.$ & $193.\pm19.$ & $880.\pm44.$ \\
$1.25 - 1.5$ & $109.\pm29.$ & $76.\pm15.$  & $131.\pm22.$ & $135.\pm20.$ & $115.\pm15.$ & $154.\pm18.$ & $720.\pm50.$ \\
$1.5 - 1.75$ & $47.\pm15.$  & $65.\pm16.$  & $62.\pm11.$  & $57.\pm10.$  & $68.\pm11.$  & $93.\pm16.$  & $391.\pm33.$ \\
$1.75 - 2.0$ & $34.\pm11.$  & $43.\pm13.$  & $50.\pm13.$  & $76.\pm18.$  & $68.\pm15.$  & $53.\pm12.$  & $323.\pm34.$ \\
$2.0 - 2.25$ & $25.\pm11.$  & $26.9\pm8.0$ & $28.4\pm8.0$ & $47.\pm20.$  & $54.\pm15.$  & $32.4\pm7.6$ & $214.\pm30.$ \\
$2.25 - 2.5$ & $17.0\pm6.0$ & $10.4\pm5.6$ & $20.4\pm7.4$ & $21.5\pm7.3$ & $24.0\pm9.2$ & $36.\pm13.$ & $129.\pm21$ \\
      \hline\noalign{\smallskip}
  sum  & $1930.\pm130.$ & $2130.\pm110.$ & $2600.\pm100.$ & $3480.\pm110.$ & $4770.\pm290.$ & $4770.\pm290.$ & $19140.\pm390.$ \\

      \hline\noalign{\smallskip}
   \end{tabular}
  \end{center}
\end{table*}

\end{document}